\documentclass[10pt,journal,twocolumn]{IEEEtran}
\usepackage{subfigure}
\usepackage{a0size}
\usepackage{amssymb}
\usepackage{multicol}
\usepackage[english]{babel}
\usepackage{epsfig}
\usepackage{bm}
\usepackage{amsfonts,color,amsthm,amsmath, lscape,graphics}
\usepackage{url}
\usepackage{algorithm}
\usepackage{algpseudocode}

\usepackage{epstopdf}
\usepackage{algorithm}
\usepackage{algpseudocode}
\usepackage{cite}
\usepackage{relsize}
\usepackage[colorlinks, linkcolor=color1, anchorcolor=blue, citecolor=color1]{hyperref}
\DeclareFontFamily{U}{matha}{\hyphenchar\font45}
\DeclareFontShape{U}{matha}{m}{n}{
      <5> <6> <7> <8> <9> <10> gen * matha
      <10.95> matha10 <12> <14.4> <17.28> <20.74> <24.88> matha12
      }{}
\DeclareSymbolFont{matha}{U}{matha}{m}{n}
\DeclareMathSymbol{\odiv}         {2}{matha}{"63}

\usepackage{fix2col}
\DeclareMathOperator*{\Maximize}{maximize}

\DeclareMathOperator*{\subj}{subject~to}

\renewcommand{\figurename}{Fig.}
\addto\captionsenglish{\renewcommand{\figurename}{Fig.}}

\definecolor{color1}{RGB}{128,0,0}

\renewcommand{\frac}{\dfrac}

\newcommand{\diag}{{\mbox{diag}}}

\definecolor{myOrange}{rgb}{1,0.5,0}
\definecolor{myGreen}{rgb}{0,0.5,0}
\definecolor{tx}{rgb}{0.1254901, 0.2196078, 0.392156}
\definecolor{rx}{rgb}{0.0, 0.5, 0.0}

\begin{document}
\title{Active Sensing for Two-Sided Beam Alignment and Reflection Design Using Ping-Pong Pilots}

\author{Tao Jiang,~\IEEEmembership{Graduate Student Member,~IEEE,}
    {Foad Sohrabi},~\IEEEmembership{Member,~IEEE,}
    and Wei Yu,~\IEEEmembership{Fellow,~IEEE}
\thanks{
Manuscript submitted to IEEE Journal on Selected Areas in Information Theory on 7 October 2022; revised on 8 March 2023; accepted on 9 May 2023. This work was supported by Huawei Technologies Canada. The materials in this work have been presented in part at Asilomar Conference on Signals, Systems
and Computers, Pacific Grove, CA, U.S.A., November 2022 \cite{asilomar_tao}. 
Tao Jiang and Wei Yu are with The Edward S.\ Rogers Sr.\ Department of
Electrical and Computer Engineering, University of Toronto, Canada. Foad Sohrabi was with The Edward S.\ Rogers Sr.\ Department of
Electrical and Computer Engineering, University of Toronto, Canada. He is now with Nokia Bell Labs, New Jersey, USA. (e-mails: taoca.jiang@mail.utoronto.ca, foad.sohrabi@gmail.com, weiyu@ece.utoronto.ca). 
The source code for this paper is available at: \url{https://github.com/taojiang-github/Active-Sensing-Beam-Alignment}
}
}
\maketitle

\begin{abstract}
Beam alignment is an important task for millimeter-wave (mmWave) communication,
because constructing aligned narrow beams both at the transmitter (Tx) and the
receiver (Rx) is crucial in terms of compensating the significant path loss in very
high-frequency bands.  However, beam alignment is also a highly nontrivial task because large antenna arrays typically have a limited number of radio-frequency chains, allowing only low-dimensional measurements of the high-dimensional channel.
This paper considers a two-sided beam alignment problem based on an alternating
ping-pong pilot scheme between Tx and Rx over multiple rounds without explicit
feedback.  We propose a deep active sensing framework in which two long
short-term memory (LSTM) based neural networks are employed to learn the
adaptive sensing strategies (i.e., measurement vectors) and to produce the 
final aligned beamformers at both sides. In the proposed ping-pong protocol, 
the Tx and the Rx alternately send pilots so that both sides can leverage local
observations to sequentially design their respective sensing and
data transmission beamformers. The proposed strategy can be
extended to scenarios with a reconfigurable intelligent surface (RIS) for designing, in addition, the reflection coefficients at the RIS for both sensing and communications.  Numerical
experiments demonstrate significant and interpretable performance improvement.
The proposed strategy works well even for the challenging multipath channel
environments.  
\end{abstract}

\section{Introduction}
\label{sec:intro}
Millimeter-wave (mmWave) is envisioned to be a key enabler for high data rate
transmission in the next generation of communication systems \cite{6515173,6736752}.
In mmWave systems, it is crucial to deploy large antenna arrays in order to focus 
narrow beams to compensate for the order-of-magnitude higher signal attenuation 
at high frequencies. However, due to the cost and energy consumption constraints,
the mmWave transceivers are generally equipped with only a limited number of
radio-frequency (RF) chains \cite{7389996}, which means only low-dimensional
signals can be observed. This makes it difficult to estimate the high-dimensional 
channel and to find the aligned beams for data transmission.
Moreover, mmWave systems are also vulnerable to signal blockage. To address this 
issue, one of the promising solutions is to use a reconfigurable intelligent surface (RIS), because of the ability of the RIS to enhance the wireless environment by
creating a reflection path \cite{9326394,huang2020holographic,wang2022beam}.
But the integration of the RIS makes pilot training even more challenging \cite{9427148}. 

This paper aims to alleviate the pilot training overhead for finding the
aligned beamformers at the transmitter (Tx) and the receiver (Rx) to maximize
the signal-to-noise ratio (SNR) of a mmWave communication link with a limited
number of RF chains. Specifically, this paper considers the two-sided beam 
alignment problem in which two multi-antenna transceivers with single RF chains need to align their 
beams together in order to find the optimal beamforming direction based on a
limited number of low-dimensional measurements designed by the two transceivers. 
Moreover, we also study the scenario in which an RIS is deployed to tackle the
blockage issue in the mmWave system. In this case, the beamformers at both
sides and the RIS coefficients are jointly optimized to maximize the overall SNR.

The two-sided beam alignment problem is highly nontrivial for several reasons.
First, the problem involves sensing a high-dimensional channel via the
low-dimensional observations through the limited number of RF chains both at the Tx 
and Rx. The design of such sensing vectors is not an easy problem to solve
analytically.  Second, because the pilots are transmitted over multiple stages,
the sensing vectors can be designed as functions of the observations in the
previous rounds.  Such an \emph{active} sensing strategy can significantly
improve the eventual beamforming gain \cite{8792366,9448070,sohrabi2021active},
but the optimal design of the sensing strategy involves sequential
exploration of the channel landscape and is extremely challenging.
Third, this paper tackles the two-sided beam alignment problem with the Tx and
the Rx actively designing their beamformers at the same time. 
Unlike the one-sided scenario \cite{8792366,8750903,9448070,sohrabi2021active}, 
the two-sided beam alignment involves the joint design of the sensing strategies at the transceivers, and hence, would normally require feedback communications between the transceivers. 
But coordination and feedback are not easy to realize before the beam alignment is achieved. 

This paper proposes a novel deep active sensing framework to tackle the
two-sided beam alignment problem for a mmWave communication link. The proposed
framework involves a ping-pong pilot transmission scheme, in which the Tx and
the Rx alternately transmit and receive pilots through the beamformers
designed by their respective active sensing units based on the pilots received
so far.  The use of the ping-pong pilot strategy allows the two sides to implicitly coordinate with each other in their respective design of the sensing strategies and eliminates the need for explicitly communicating the designed beamformers from the Rx to the Tx.

To account for the sequential nature of the learning task, this paper proposes to utilize a recurrent neural network (RNN) with long short-term memory (LSTM) as the active sensing units at both the Tx and RX to efficiently extract the relevant information from the
received pilots over multiple stages to design the Tx and Rx beamformers both
for the sensing phase and for the final data
transmission phase. The proposed active sensing strategy can be
readily extended to the RIS-assisted mmWave system, where the beamformers and
reflection coefficients are jointly and actively designed.  Overall, the
proposed design is shown to significantly reduce the pilot overhead and to
enhance the overall SNR for the mmWave link, while producing interpretable beamforming and reflection patterns.

\subsection{Related Works}
The two-sided beam alignment problem for mmWave systems has been studied in \cite{6847111,8356247,8625694,8962355,9457655,7996580,8644444,9154340}. To reduce the pilot training overhead, most of the nonadaptive beam alignment algorithms leverage the compressive sensing technique to find the angle of arrivals (AoAs) and angle of departures (AoDs) \cite{6847111,8356247,8625694,8962355}, where the sparsity of the mmWave channel is exploited in designing the AoA/AoD recovery algorithm, but the sensing beamformers are typically randomly generated to explore all the possible directions. To reduce the pilot training overhead further, iterative search algorithms have been proposed to actively design the sensing beamformers based on historical observations. For instance, the works of \cite{6847111,7996580,8644444} propose to construct sensing beamformers adaptively according to a hierarchical codebook, from which the next sensing beamformers are chosen, based on the power of the received pilots. Specifically, \cite{8644444} proposes a scheme where upon receiving a pilot signal, the transmitter/receiver would sort the received signal power of all the pilots received so far, and then split the beam into narrower beams in the direction with the strongest received power.  However, such codebook-based schemes are not necessarily the optimal strategy for designing the sensing beamformers, especially at low SNR. 

A better design of the sensing strategy is to
select the next sensing beamformers based on the current posterior distribution
of the AoA/AoD. Such posterior matching based methods have been shown to have
superior performance for the one-sided beam alignment problem \cite{8792366},
and are extended to the two-sided scenarios in \cite{9457655, 9154340}.
However, the posterior matching based methods are mostly suitable only for
simple channel models, such as the single-path channel for which the posterior
distribution can be readily computed. For the multipath channel,
calculating the posterior distribution of the AoA/AoD after each new
observation becomes computationally infeasible. Moreover, most of the existing works
are based on hierarchical codebooks, which reduce the search complexity
but can significantly degrade the final AoA/AoD estimation performance.

To eliminate the codebook constraint on designing the active sensing strategy,  \cite{9448070} proposes to map the posterior distribution to the next sensing vector by a fully connected deep neural network (DNN) and demonstrates significant performance improvement over the codebook based posterior matching method \cite{8792366}. A similar work \cite{9473733} also takes the posterior distribution as input to a DNN to design the next beam probing direction. But computing the posterior distribution is feasible only for the single-path channel model. Instead of using the posterior distribution as input to design the sensing vectors, a follow-up paper \cite{sohrabi2021active} proposes a deep learning framework based on the LSTM neural network architecture to summarize the information contained in the observations automatically, which further improves the one-sided beam alignment performance and is applicable to multipath channel models. 
However, the beam alignment method in \cite{sohrabi2021active} is not directly applicable to the two-sided scenario for the following reasons. First, the two scenarios have different solution structures. The optimal two-sided data transmission beamformers need to be designed to match the left and right singular vectors corresponding to the largest singular value of the channel matrix, while the one-sided beam alignment only needs to match a single channel vector. Moreover, for the two-sided case, the design of sensing beamformers at both sides needs to search over the singular vector pair of the channel matrix, which is a much larger search space than the space of sensing vectors in the one-sided case. Specifically, the two-sided scenario requires tackling the challenging problem of the joint design of sensing strategies at both transmitter and receiver sides which implicitly cooperate with each other \emph{over the air}, while the one-sided scenario designs the sensing vectors independently from just one side. Thus, the generalization of the method in \cite{sohrabi2021active} for two-sided beam alignment is highly nontrivial.

We mention here that the beam alignment problem has also been investigated from the deep learning perspective in \cite{8395149,8842625,8999545,9425522,attiah2021deep}, but these works are all limited to the one-sided beam alignment problem. We remark that this paper considers a beam alignment problem within a coherence block where the Tx and Rx are not moving and the channel between the Tx and the Rx remains constant, while \cite{9481259,9819358,han2023} address the beam tracking problem in the context of a time-varying channel model where the Rx moves following some Markov chain.

This paper also investigates the use of 
RIS for providing a controllable and reliable reflection link for the mmWave channel \cite{wang2022beam}, which is vulnerable to signal blockage. The resulting beam and reflection alignment problem has been investigated previously in \cite{9382000,9129778,alexandropoulos2022near,9556614}. For example, by exploiting the sparsity of the mmWave channel, a compressed sensing based channel estimation method is proposed in \cite{9382000}, where the codebook for RIS coefficients is optimized based on the location information to reduce the pilot training cost. In \cite{9129778}, the authors propose a multi-beam training method to find the optimal RIS beam direction while assuming that the base station (BS) has already aligned its beam to the line-of-sight direction of the RIS. For RIS beam training, \cite{alexandropoulos2022near} proposes a variable-width hierarchical phase shift codebook in order to account for the near field propagation environment of the RIS. The paper \cite{9556614} considers a more complicated scenario where the beamformers at the BS and the RIS reflection coefficients are jointly designed in the beam training phase to serve a single-antenna user. None of the existing works treat the more general case of joint two-sided beam alignment and RIS reflection design to maximize the SNR of the overall communication link, which the current paper seeks to tackle.

\subsection{Main Contributions}
The paper proposes an active sensing framework for the two-sided beam alignment 
and RIS reflection coefficients design problem using a ping-pong pilot scheme. 
Our main contributions are as follows:

\subsubsection{Ping-Pong Pilot Scheme} This paper proposes an active sensing
framework based on a ping-pong pilot training scheme, in which the Tx and the Rx
alternately transmit and receive pilots so that all the sensing beamformers
can be designed based on local observations. In particular, in each round, the
Tx first uses its transmit beamformer to send a pilot, which the Rx receives
with a receive beamformer; the Rx then uses a transmit beamformer to send back
a pilot, which the Tx receives with a receive beamformer. All these
transmit/receive beamformers at each side are adaptively designed by the
proposed active sensing unit based on locally received historical pilots. 

The proposed ping-pong pilot scheme has the advantage that unlike most
existing works for two-sided beam alignment \cite{6847111,9154340}, it implicitly enables coordination between the two sides without requiring error-free feedback between the Tx and the Rx. 
Such feedback would lead to extra delay and communications costs
\cite{khalili2022optimal}. More importantly, it is challenging to establish 
an error-free link at the initialization stage, because the SNR between the Tx 
and Rx is very low in mmWave systems prior to beam alignment. 

We note that the ping-pong beam training has been used for fully digital MIMO systems \cite{895022,1323251,7575663} and recently been extended to the hybrid beamforming scenarios \cite{7996580,8644444}. Unlike \cite{7996580,8644444}, where the next sensing beamformers are designed based on the power of the received pilots and a predefined codebook, the proposed active sensing method is codebook-free and can exploit both the phase and magnitude information of the received pilots and can automatically take into account the channel and noise distributions.

\subsubsection{Active Sensing Strategy using LSTM  Network}
The proposed deep active sensing unit consists of two LSTMs deployed
separately at the Tx and the Rx. Each LSTM takes the locally received pilot as input
and outputs the transmit/receive sensing beamformers (i.e., the measurement vectors) for the next round of
pilot training. In the proposed deep active sensing unit, the LSTM cell learns
to update its hidden state and cell state using the newly received pilot
symbol. The hidden state information vector
can efficiently summarize sufficient information for designing the successive
sensing vectors. The hidden state at each side is mapped to the
next transmit/receive sensing beamformers via two different fully connected
DNNs. The final cell state at each side is mapped to the beamformers for data
transmission via another DNN. In this way, the proposed active sensing
framework is codebook-free and is suitable for a generic multipath channel model
without the need for calculating the posterior distribution.

The active sensing unit described above corresponds to one round of the ping-pong pilot training. By concatenating the active sensing units for all the transmission rounds together, we train the overall deep learning framework end-to-end to maximize the beamforming gain so that the adaptive sensing strategies of the transceivers can be jointly optimized. We remark that the joint design of adaptive sensing strategies for two-sided beam alignment is a highly challenging task.

\subsubsection{Beam Alignment with RIS}

The proposed deep active sensing framework is further applied to the
scenario in which an RIS is placed between the Tx and the Rx to tackle the blockage
issue. In this scenario, the reflection coefficients are adaptively designed
using additional DNNs at an RIS controller, which has access to the hidden state and cell state
information from both the Tx and the Rx. In particular, the next reflection coefficients
for the pilot transmissions in both directions are designed by two different
DNNs that take the latest hidden states from both sides as inputs. Another DNN
is used at the final stage to map the cell states from both sides to the
reflection coefficients in the data transmission phase.

\subsubsection{Interpretability}
The proposed active sensing approach is shown to have superior performance 
as compared to the conventional methods. It works well even for the challenging
multipath mmWave environments. More importantly, we show that the sensing
beamforming patterns learned by the proposed active sensing units have strong
interpretability. The active sensing strategy has learned to produce
beamforming and reflection patterns that match the strongest direction of
the channel most of the time, through a sequential exploration of the channel
landscape over multiple stages.

\subsection{Paper Organization and Notations}
The remaining part of this paper is organized as follows. Section \ref{sec:sys_two_sided}
describes the mmWave system model and the problem formulation. Section \ref{sec:proposed_LSTM} presents the
proposed active sensing framework. Section \ref{sec:sim_hybrid} presents simulation results for the mmWave system.
Section \ref{sec:sys_ris} extends the proposed active sensing framework to the RIS-assisted system.  Section
\ref{sec:conclusion} concludes the paper.

The notations used in this paper are as follows. We use lower-case letters to
denote scalars, and lower-case and upper-case boldface letters to denote
vectors and matrices, respectively, e.g., $\bm v\in \mathbb{C}^N$ is a
complex vector of dimension $N$, $\bm A\in\mathbb{C}^{M\times N} $ is a
$M\times N$ complex matrix. For a vector $\bm v$, $[\bm v]_n$ denotes its $n$-th entry. For a matrix $\bm A$, $\bm A^\top$ and $\bm A^{\sf
H}$ denote its transpose and conjugate transpose,
respectively. We use $\Re(\cdot)$ and $\Im(\cdot)$ to denote the real and imaginary parts of the argument. We use $|v|$ to denote the magnitude of the complex scalar $v$. We use $\circ$ to denote element-wise product.
\section{Two-Sided Beam Alignment}
\label{sec:sys_two_sided}

\subsection{System Model}
\begin{figure}[t]
\centering
    \includegraphics[width=7cm]{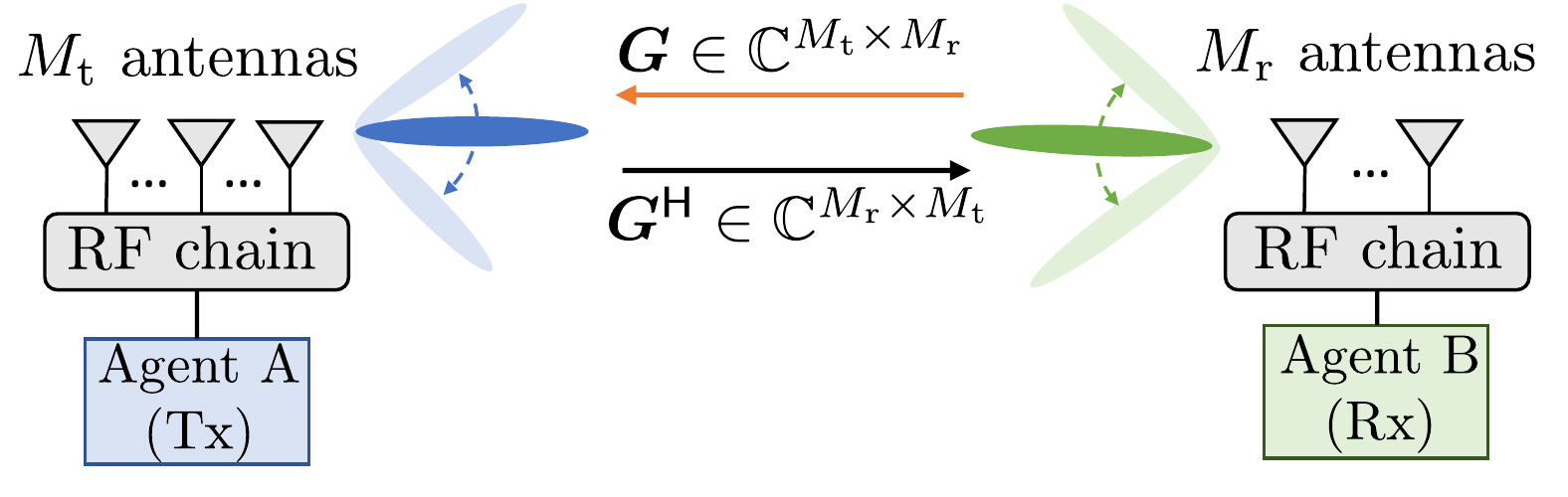}
    \caption{Two-sided beam alignment problem in mmWave system.}
    \label{fig:system_model_hybrid}
\end{figure}
Consider a mmWave MIMO communication system consisting of a Tx, also called agent A, with $M_{\rm t}$ antennas configured as a uniform linear array (ULA), and a Rx, also called agent B, with $M_{\rm r}$ antennas as a ULA. The Tx and the Rx each have a single RF chain. Let $\bm w_{\rm t}\in\mathbb{C}^{M_{\rm t}}$ and $\bm w_{\rm r}\in\mathbb{C}^{M_{\rm r}}$ denote the beamforming vectors at the Tx and the Rx side, respectively. Without loss of generality, the transmitted signal is subject to a power constraint and the beamformers are normalized\footnote{This can be implemented by adding an amplifier to the phase-shifter at each antenna element \cite{8750903}. We remark that our proposed method can be readily applied also to the case where the beamformers have constant modulus constraints, i.e., each antenna element can only change the phase of the signal.}  such that $\|\bm w_{\rm t}\|_2=\|\bm w_{\rm r}\|_2=1$. To establish a reliable link between the Tx and the Rx, the beamforming vectors $\{\bm w_{\rm t}, \bm w_{\rm r}\}$ should be jointly optimized according to the channel state information (CSI) so that the achievable rate (or equivalently the SNR) of the communication link is maximized. This is called the two-sided beam alignment problem. 

As shown in Fig.~\ref{fig:system_model_hybrid}, let matrix $\bm G\in\mathbb{C}^{M_{\rm t} \times M_{\rm r}}$ denote the uplink channel matrix from the Rx to the Tx, then the downlink channel is denoted by $\bm G^{\sf H}$, where we assume the system operates in the time-division duplex (TDD) mode with channel reciprocity. We assume a block fading channel model, where the channel remains constant over a coherence interval but changes independently over different coherence intervals. To capture the intrinsic sparse nature of the  mmWave propagation environments, a mmWave channel is typically modeled by a sparse multipath channel as follows \cite{7389996}:
\begin{align}
    \bm G = \sum_{i=1}^{L_{\rm p}}\alpha_i \bm a_{\rm t}(\phi_{\rm t}^i)\bm a_{\rm r}^{\sf H}(\phi_{\rm r}^i),
\end{align}
where $L_{\rm p}$ denotes the number of paths, $\alpha_i\sim\mathcal{CN}(0,1)$ denotes the flat fading coefficient of the $i$-th path, $\phi_{\rm t}^i$ denotes the AoA of the $i$-th path to Tx, and $\phi_{\rm r}^i$ denotes the AoD of the $i$-th path from Rx, and $\bm a_{\rm t}(\cdot)$ and $\bm a_{\rm r}(\cdot)$ are the steering vectors as given by the following (assuming half-wavelength antenna spacing):
\begin{subequations}
    \begin{align}
        &[\bm a_{\rm t}(\phi_{\rm t}^i)]_m = e^{j\pi(m-1)\sin(\phi_{\rm t}^i)}, ~~m=1,\cdots,M_{\rm t},\\
        &[\bm a_{\rm r}(\phi_{\rm r}^i)]_m = e^{j\pi(m-1)\sin(\phi_{\rm r}^i)}, ~~m=1,\cdots,M_{\rm r}.
    \end{align}
\end{subequations}
We assume a narrowband channel model in this paper. The generalization of the proposed approach to a wideband scenario, e.g., OFDM, in which a common beamformer is applied across multiple frequency bands, is left as future work.

Let $x\in\mathbb{C}$ with $\mathbb{E}[|x|^2]=P$ denote the data symbol, then the received signal at the Rx can be expressed as 
\begin{equation}\label{eq:received}
    r = \bm w_{\rm r}^{\sf H}\bm G^{\sf H}\bm w_{\rm t}x+n,
\end{equation}
where $n\sim\mathcal{CN}(0,\sigma_0^2)$ is the effective additive Gaussian noise. In order to maximize the transmission rate, the beamforming vectors should be designed to maximize the beamforming gain $|\bm w_{\rm r}^{\sf H}\bm G^{\sf H}\bm w_{\rm t}|^2$.

Given perfect CSI $\bm G$, the beamforming vectors $\{\bm w_{\rm t}^\star, \bm w_{\rm r}^\star \}$ that maximize the beamforming gain are given by 
\begin{subequations}\label{eq:optimal_bf}
    \begin{align}
        &\bm{w}_{\rm t}^\star  =  \bm u_{\max}/{\|\bm u_{\max}\|_2},\\
        &\bm{w}_{\rm r}^\star  = \bm v_{\max}/{\| \bm v_{\max}\|_2},
    \end{align}
\end{subequations}  
where $\bm u_{\max}$ and $\bm v_{\max}$ are respectively the left and the right singular vectors associated with the largest singular value of the matrix $\bm G$. That is, the beamformers at the Tx and the Rx should be respectively aligned with the singular vectors $\bm u_{\max}$ and $\bm v_{\max}$ of the channel matrix $\bm G$. 

However, the channel matrix $\bm G$ is not known initially. The transceivers
need to effectively obtain the CSI from a pilot training phase. 
The problem of estimating the top singular vector pair in \eqref{eq:optimal_bf} has been investigated in a fully digital MIMO setup \cite{1323251}. In particular, \cite{1323251} proposes an approach where pilots are sent alternately between Tx and Rx to imitate the power iteration method for eigen-decomposition of a matrix. In the absence of noise, this algorithm can be shown to converge to the top singular vectors pair with linear convergence rate \cite{golub2013matrix}. This suggests the possibility of significant reductions in pilot overhead, compared to estimating the entire channel matrix $\bm G$. However, the power iteration method involves observing the result of a matrix-vector multiplication, so it requires the number of RF chains to be equal to the number of antennas, which is a fundamentally different setup from the current scenario where the observation is a single complex number as in \eqref{eq:received}.  Additionally, the approach based on the power iteration method is very sensitive to noise and is only applicable in high SNR scenarios.

A conventional strategy for channel estimation involves sending pilots in the uplink and estimating the channel matrix at the Tx. The transmit and receive sensing beamformers in the
pilot training phase (i.e., the measurement vectors) are typically randomly
chosen so that all directions are explored \cite{8356247,8625694}. 
Adaptive pilot training schemes have been proposed in \cite{7996580,8644444}
to reduce the pilot training overhead, but the existing works typically choose
the sensing beamformers from a codebook, which is restrictive. Further, many existing algorithms depend on posterior distribution calculation \cite{9457655, 9154340}, 
which is computationally feasible only for the single-path channel model. Moreover, most of the existing
adaptive sensing strategies require error-free feedback between the Tx and the Rx
\cite{9457655, 6847111} during the pilot training phase, which adds to the
communication cost and is difficult to realize, because the SNR of the
communication link is usually quite low until beam alignment is achieved.

The key mathematical difficulty in designing an optimal adaptive sensing strategy for
beam alignment is the complexity of the optimization problem over the sensing 
beamformers across multiple stages. Conventional approaches based on mathematical
models of the underlying channel inevitably lead to intractable problem formulations.
This paper proposes instead 
to bypass channel modeling and to use a deep learning approach to adaptively 
design the sensing beamformers at both sides during pilot training. Moreover, the subsequent
beamformers for data transmission are also learned in an end-to-end fashion. 
As shown in Section~\ref{subsect:sim_hybrid} and \ref{sec:simulation_results}, the proposed method
can efficiently perform beam alignment even in the short pilot-length regime. 
This is all accomplished based on a novel ping-pong pilot protocol without 
explicit feedback. 

\subsection{Ping-Pong Pilot Training Protocol}\label{subect:ping-ping}
\begin{figure}
\centering
    \includegraphics[width=8.6cm]{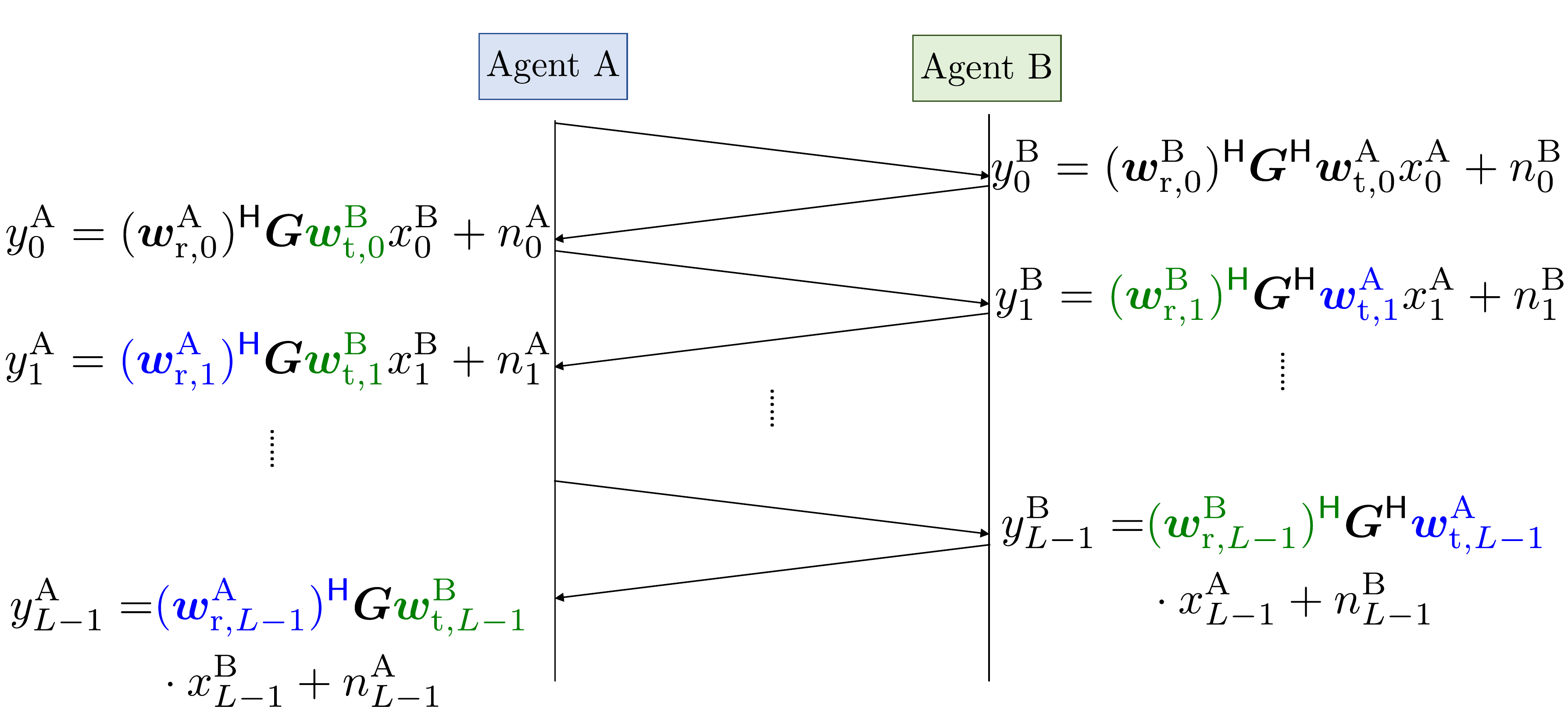}
    \caption{Proposed ping-pong pilot training protocol. The sensing beamformers actively designed at agent A and agent B are highlighted as blue (e.g., \textcolor{blue}{$\bm{w}^{\rm A}_{{\rm t},1},~\bm{w}^{\rm A}_{{\rm r},1}$}) and green (e.g., \textcolor{rx}{$\bm{w}^{\rm B}_{{\rm t},0},~\bm{w}^{\rm B}_{{\rm r},1}$}), respectively. 
    The initial vectors $\bm w^{\rm A}_{\rm t,0}$, $\bm w^{\rm A}_{\rm r,0}$ and $\bm w^{\rm B}_{\rm r,0}$ are fixed and can be learned from the channel statistics.}
    \label{fig:pilot_transmission_pingpong1}
\end{figure}
The ping-pong pilot training protocol is illustrated in Fig.~\ref{fig:pilot_transmission_pingpong1}, in which the pilot symbols are sent back and forth between agent A (i.e., Tx) and agent B (i.e., Rx) such that each side gathers the required information to design its own beamformer for the data transmission phase. 

In the $\ell$-th transmission round, agent A first sends a pilot symbol $x^{\rm A}_\ell$ under a power constraint $\mathbb{E}[|x^{\rm A}_\ell|^2]\le P_1$ to agent B, then the received pilot symbol at agent B is given by 
\begin{align}
    y^{\rm B}_\ell = (\bm w^{\rm B}_{{\rm r},\ell})^{\sf H} \bm G^{\sf H}\bm w^{\rm A}_{{\rm t},\ell} x^{\rm A}_\ell+n^{\rm B}_\ell,\quad\ell=0,\cdots,L-1,
\end{align}
where the vectors $\bm w^{\rm A}_{{\rm t},\ell}\in\mathbb{C}^{M_{\rm t}}$ and $\bm w^{\rm B}_{{\rm r},\ell}\in\mathbb{C}^{M_{\rm r}}$ are respectively the transmit sensing beamforming vector at agent A and the receive sensing beamforming vector at agent B in the $\ell$-th round of pilot transmission, and $n^{\rm B}_\ell\sim\mathcal{CN}(0,\sigma^2)$ is the effective additive Gaussian noise. The beamforming vectors in the pilot training phase are called sensing vectors to distinguish them from the beamforming vectors in the data transmission phase. After receiving the pilot, as shown in Fig.~\ref{fig:pilot_transmission_pingpong1}, agent B sends back a pilot symbol ${x}^{\rm B}_\ell$ under a power constraint $\mathbb{E}[|{x}^{\rm B}_\ell|^2]\le P_2$ to agent A. Similarly, the received pilot at agent A is given by 
\begin{align}
    {y}^{\rm A}_\ell =({\bm{w}}^{\rm A}_{{\rm r},\ell})^{\sf H} {\bm G} {\bm{w}}^{\rm B}_{{\rm t},\ell} {x}^{\rm B}_\ell+{n}^{\rm A}_\ell,\quad\ell=0,\cdots,L-1,
\end{align}
where the vectors ${\bm{w}}^{\rm A}_{{\rm r},\ell}\in\mathbb{C}^{M_{\rm t}}$ and ${\bm{w}}^{\rm B}_{{\rm t},\ell}\in\mathbb{C}^{M_{\rm r}}$ are the sensing vectors at agent A and agent B, respectively, and ${n}^{\rm A}_\ell\sim\mathcal{CN}(0,\sigma^2)$ is the effective additive Gaussian noise.  Without loss of generality, we can set ${x}_\ell^{\rm A} = \sqrt{P_1}$ and ${x}^{\rm B}_\ell = \sqrt{P_2}$. After $L$ rounds of pilot transmission, each of the transceivers obtains $L$ measurements of the channel, which can be utilized to design their own beamforming vector for the data transmission phase. The overall pilot training overhead is $2L$ for $L$ rounds of pilot transmission.

Unlike the conventional schemes in which both the transmit and receive beamformers are designed at one node, the presented protocol designs the beamformers locally, hence it does not require an explicit feedback procedure.

\subsection{Active Sensing for Two-Sided Beam Alignment}
The proposed active sensing approach is an adaptive sensing method because each sensing (or measurement) vector is designed based on the previously received pilot symbols.  At the beginning of the $\ell$-th ping-pong round, as shown in Fig.~\ref{fig:pilot_transmission_pingpong1}, agent A sends a pilot to agent B. After receiving the observation $y_{\ell}^{\rm B}$, agent $B$ utilizes all the historical observations $\{y^{\rm B}_{i}\}_{i=0}^{\ell} $ to design its next transmit sensing beamformer ${\bm w}^{\rm B}_{{\rm t},\ell}$, as well as the receive sensing beamformer $\bm w^{\rm B}_{{\rm r},\ell+1}$ in the next round, i.e., 
\begin{subequations}\label{eq:pilot_rx}
    \begin{align}
    &{\bm w}^{\rm B}_{{\rm t},\ell} = f_{\rm t,\ell}^{\rm B}\left(\{y^{\rm B}_{i}\}_{i=0}^{\ell}\right), ~ \ell=0,\cdots,L-1,\\
    &\bm w^{\rm B}_{{\rm r},\ell+1} = {f}_{\rm r,\ell}^{\rm B}\left(\{y^{\rm B}_{i}\}_{i=0}^{\ell}\right), ~\ell=0,\cdots,L-2,
    \end{align}
\end{subequations}
where $f_{\rm t,\ell}^{\rm B}: \mathbb{C}^{\ell+1}\rightarrow\mathbb{C}^{M_{\rm r}}$ and ${f}_{\rm r,\ell}^{\rm B}: \mathbb{C}^{\ell+1}\rightarrow\mathbb{C}^{M_{\rm r}}$ are the corresponding active sensing schemes at agent B. The outputs of the functions $f_{\rm t,\ell}^{\rm B}$ and ${f}_{\rm r,\ell}^{\rm B}$ should also satisfy the unit $\ell_2$-norm constraints, i.e., $\|{\bm w}^{\rm B}_{{\rm t},\ell}\|_2=\|\bm w^{\rm B}_{{\rm r},\ell+1}\|_2=1$.

At the end of the $\ell$-th ping-pong round, agent A  has made $\ell+1$ observations, i.e., $\{y^{\rm A}_i\}_{i=0}^{\ell}$, so its next transmit sensing beamforming vector $\bm w^{\rm A}_{{\rm t}, \ell+1}$ and next receive sensing beamforming vector ${\bm w}^{\rm A}_{{\rm r},\ell+1}$ in the next round can be designed as functions of these observations, i.e., 
\begin{subequations}\label{eq:pilot_tx}
    \begin{align}
    &\bm w^{\rm A}_{{\rm t}, \ell+1} = f^{\rm A}_{\rm t,\ell}\left(\{y^{\rm A}_i\}_{i=0}^{\ell}\right), ~ \ell=0,\cdots,L-2,\\
    &{\bm w}^{\rm A}_{{\rm r},\ell+1} = {f}^{\rm A}_{\rm r,\ell}\left(\{y^{\rm A}_i\}_{i=0}^{\ell} \right),  ~ \ell=0,\cdots, L-2,
    \end{align}
\end{subequations}
where $f^{\rm A}_{\rm t,\ell}: \mathbb{C}^{\ell+1}\rightarrow\mathbb{C}^{M_{\rm t}}$ and $f^{\rm A}_{\rm r,\ell}: \mathbb{C}^{\ell+1}\rightarrow\mathbb{C}^{M_{\rm t}}$ are respectively the transmit and receive active sensing strategies that map the historical observations to the sensing vectors in the next round, while satisfying the unit $\ell_2$-norm constraints, $\|\bm w^{\rm A}_{{\rm t}, \ell+1}\|_2=\|{\bm w}^{\rm A}_{{\rm r},\ell+1}\|_2=1$. 
In the initial stage, since no prior observation is available, we fix the initial beamformers $\bm w^{\rm A}_{{\rm t},0}$, $\bm w^{\rm B}_{{\rm r},0}$, and $\bm w^{\rm A}_{{\rm r},0}$ to be some fixed unit-norm vectors, which can be designed based on channel statistics.

After $L$ rounds of pilot transmission, the two agents design their respective beamformers for data transmission based on all the historical received pilots. In particular, the final beamforming vectors in the data transmission phase are given by%
\begin{subequations}\label{eq:final_w}
    \begin{align}
    &\bm w_{\rm t} =g_{\rm t}\left(\{y^{\rm A}_i\}_{i=0}^{L-1}\right),\\
    &\bm w_{\rm r} = g_{\rm r}\left(\{y^{\rm B}_{i}\}_{i=0}^{L-1}\right),
    \end{align}
\end{subequations}
where $g_{\rm t}: \mathbb{C}^L\rightarrow\mathbb{C}^{M_{\rm t}}$ and $g_{\rm r}: \mathbb{C}^L\rightarrow\mathbb{C}^{M_{\rm r}}$  are functions that map the received pilots to the final beamforming vectors with unit-norm constraints at agent A (i.e., Tx) and agent B (i.e., Rx), respectively.

This paper aims to find the optimal active sensing strategies together with the mapping functions at the final stage so that the overall beamforming gain for data transmission is maximized. The overall problem can be formulated as
\begin{subequations}\label{eq:problem_formulation}
    \begin{align}
        &\underset{\mathcal{F}}{\Maximize}~~~ \mathbb{E}\left[|\bm w_{\rm r}^{\sf H}\bm G^{\sf H}\bm w_{\rm t}|^2\right]\\
        &\subj ~~~\eqref{eq:pilot_tx}, \eqref{eq:pilot_rx}, ~\text{and}~\eqref{eq:final_w},
    \end{align}
\end{subequations}
where the optimization variables are a set of functions
\begin{equation}
    \begin{aligned}
        \mathcal{F}=&\left\{\{f^{\rm A}_{\rm t, \ell}(\cdot)\}_{\ell=0}^{L-2},\{f^{\rm A}_{\rm r,\ell}(\cdot)\}_{\ell=0}^{L-2},
        \{f^{\rm B}_{\rm t, \ell}(\cdot)\}_{\ell=0}^{L-1},\right.\\
        &~\left. \{f^{\rm B}_{\rm r,\ell}(\cdot)\}_{\ell=0}^{L-2},
        g_{\rm t}(\cdot),g_{\rm r}(\cdot)\right\},
    \end{aligned}
\end{equation}
and the expectation is taken over all the stochastic parameters in the system, i.e., the channels and the noise. 

Solving the problem \eqref{eq:problem_formulation} is computationally challenging because the optimization variables are high-dimensional functions. Moreover, the input dimensions of the functions in $\mathcal{F}$ increase with the number of rounds, so the complexity also scales accordingly. 
The main idea of this paper is that an RNN can be used to solve this optimization problem efficiently.

\begin{figure}[t]
    \centering
    \includegraphics[width=8cm]{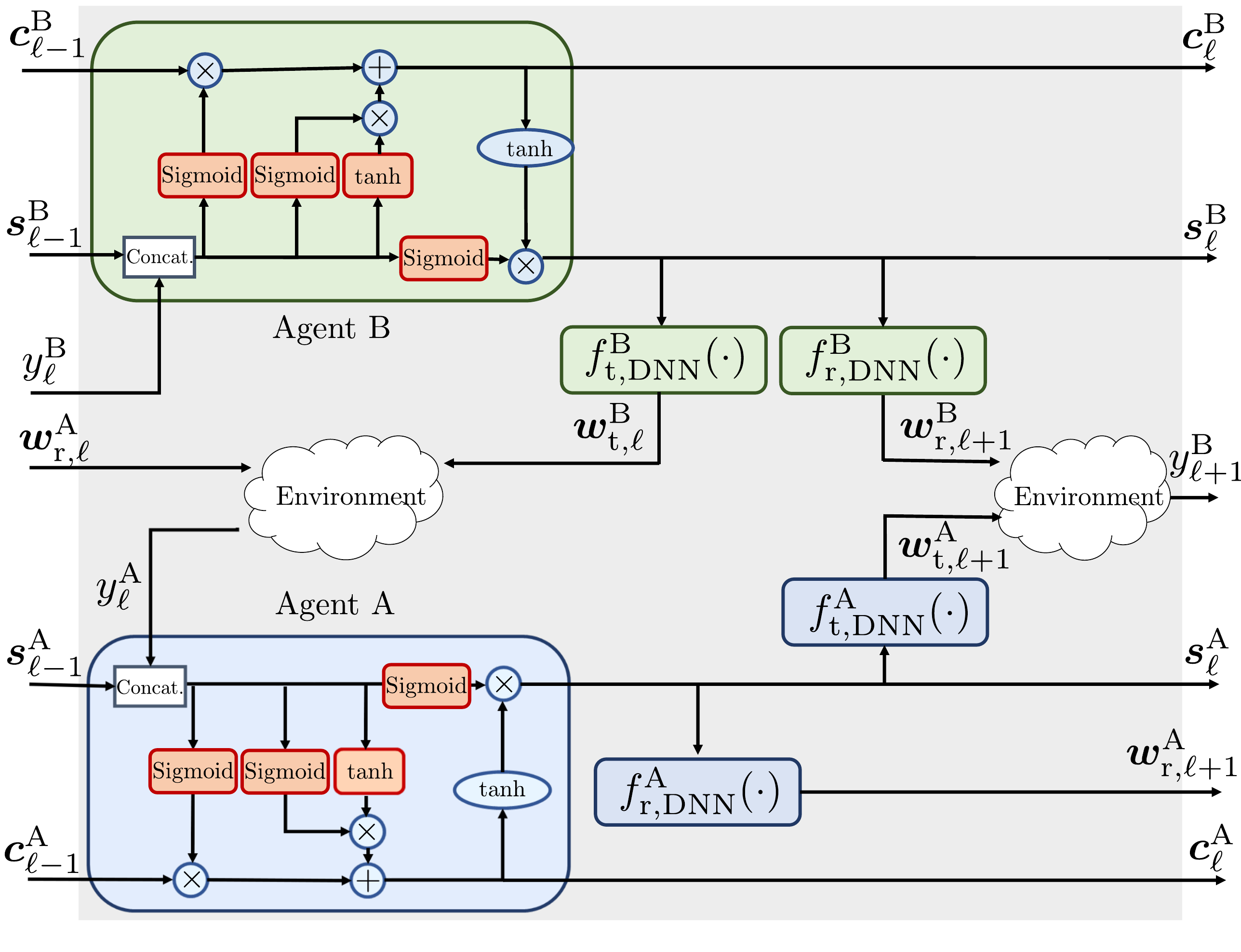}
    \caption{Proposed active sensing unit for the two-sided beam alignment problem in the $\ell$-th ping-pong pilot training round.}\label{fig:lstm_framework}
\end{figure}

\begin{figure*}[t]
    \centering
    \includegraphics[width=17cm]{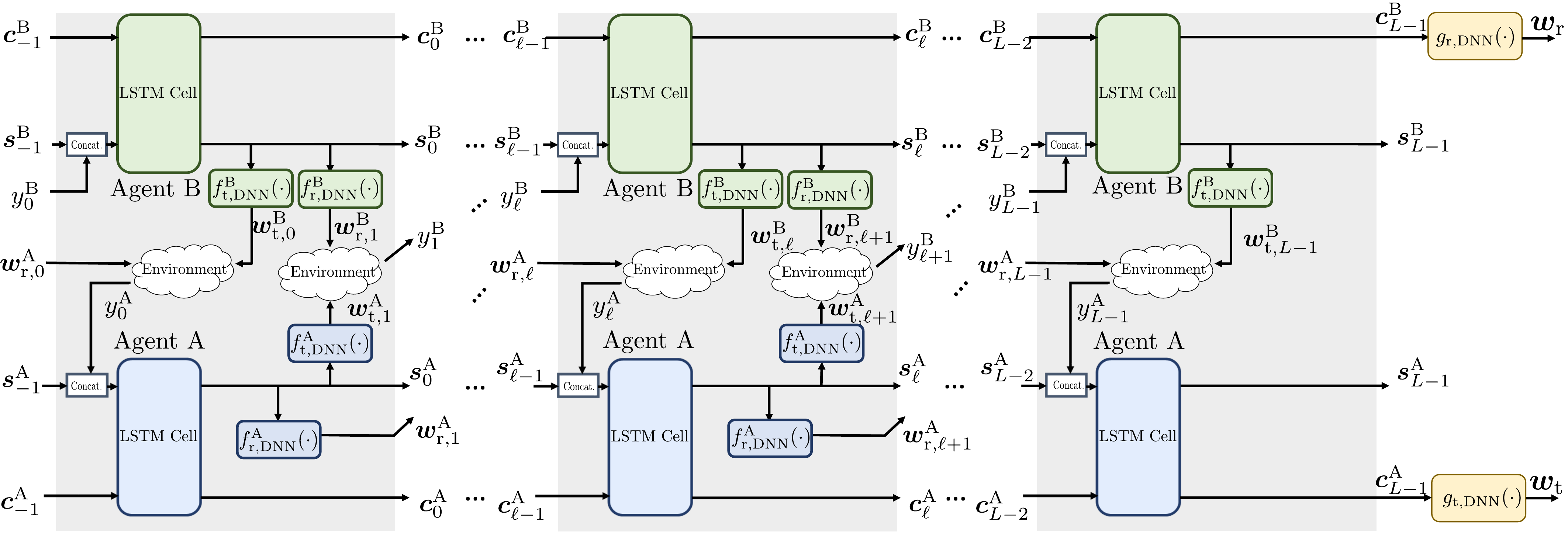}
    \caption{Proposed overall active sensing framework for the two-sided beam alignment problem with $L$ ping-pong pilot training rounds.}\label{fig:overall_lstm}
\end{figure*}

\section{Proposed Deep Learning Framework}
\label{sec:proposed_LSTM}

In this section, we propose a deep learning framework to parameterize the mapping in $\mathcal{F}$ and to solve the optimization problem (\ref{eq:problem_formulation}). In particular, the sequential nature of the active sensing problem motivates the use of RNN. 
Further, we propose to use two LSTM cells \cite{6795963}, one on each side, to automatically summarize the historical observations into fixed-dimensional state vectors, which are used for designing the subsequent sensing vectors. This is motivated by the fact that LSTM can capture the
correlations over long sequences. 
It has the ability to summarize effectively a sufficient statistic from historical observations, thereby preventing the dimension of the input from growing with the number of pilot transmission rounds. 

Fig.~\ref{fig:lstm_framework} shows the architecture of the proposed active sensing unit in the $\ell$-th pilot transmission round. In particular, the proposed active sensing unit consists of two LSTMs deployed at agent A and agent B, respectively. 

At the Rx side, given the new observation ${y}_\ell^{\rm B}$, its LSTM cell outputs the updated cell state vector ${\bm c}^{\rm B}_{\ell}$ and hidden state vector ${\bm s}^{\rm B}_{\ell}$ according to the following equations \cite{6795963}:
\begin{subequations}
    \begin{align}
    \bm{f}^{\rm B}_\ell &= \operatorname{sigmoid}(\bm{W}^{\rm B}_\text{f} \bm{y}^{\rm B}_{\ell} + \bm{U}^{\rm B}_\text{f} {\bm s}_{\ell-1}^{\rm B} + \bm{b}^{\rm B}_\text{f}),\\
    \bm{i}^{\rm B}_\ell &= \operatorname{sigmoid}(\bm{W}^{\rm B}_\text{i} \bm{y}^{\rm B}_{\ell} + \bm{U}^{\rm B}_\text{i} {\bm s}_{\ell-1}^{\rm B} + \bm{b}^{\rm B}_\text{i}),\\
    \bm{o}^{\rm B}_\ell &= \operatorname{sigmoid}(\bm{W}^{\rm B}_\text{o} \bm{y}^{\rm B}_{\ell} + \bm{U}^{\rm B}_\text{o} {\bm s}_{\ell-1}^{\rm B} + \bm{b}^{\rm B}_\text{o}),\\
    {\bm c}^{\rm B}_{\ell} &= \bm{f}^{\rm B}_\ell \circ {\bm c}^{\rm B}_{\ell-1} + \bm{i}^{\rm B}_\ell \circ \operatorname{tanh}(\bm{W}^{\rm B}_\text{c} \bm{y}^{\rm B}_{\ell} + \bm{U}^{\rm B}_\text{c} {\bm s}_{\ell-1}^{\rm B} + \bm{b}^{\rm B}_\text{c}),\\ 
    {\bm{s}}^{\rm B}_{\ell} &= \bm{o}^{\rm B}_\ell \circ \operatorname{tanh}({\bm{c}}^{\rm B}_\ell),
    \label{eq:info_state_rx}
    \end{align}
\end{subequations}
where ${\bm y}^{\rm B}_\ell = [\Re({y}^{\rm B}_\ell), \Im({y}^{\rm B}_\ell) ]^\top$ is the concatenation of real and imaginary parts of ${y}^{\rm B}_\ell$, $\{\bm{W}^{\rm B}_\text{f},\bm{W}^{\rm B}_\text{i}, \bm{W}^{\rm B}_\text{o},\bm{W}^{\rm B}_\text{c},\bm{U}^{\rm B}_\text{f},\bm{U}^{\rm B}_\text{i},\allowbreak\bm{U}^{\rm B}_\text{o},\bm{U}^{\rm B}_\text{c}\}$ and $\{\bm{b}^{\rm B}_\text{f},\bm{b}^{\rm B}_\text{i},\bm{b}^{\rm B}_\text{o},\bm{b}^{\rm B}_\text{c}\}$ are respectively the trainable weights and biases in the LSTM cell. Moreover, the forget gate's activation vector $\bm{f}_{\ell}^{\rm B}$, input/update gate's activation vector $\bm{i}_{\ell}^{\rm B}$, and output gate's activation vector $\bm{o}_{\ell}^{\rm B}$ are intermediate vectors generated within the LSTM unit to update the cell state vector $\bm{c}_{\ell}^{\rm B}$ and the hidden state vector $\bm{s}_{\ell}^{\rm B}$.

Similarly, the Tx utilizes another LSTM with the same architecture but with different trainable parameters. As can be seen from Fig.~\ref{fig:lstm_framework}, given the new observation $y^{\rm A}_\ell$ at the Tx side, the LSTM updates its cell state vector $\bm{c}^{\rm A}_{\ell}$ and hidden state vector ${\bm s}^{\rm A}_{\ell}$ according to the LSTM cell. To initialize the cell state and hidden state vectors, we set ${\bm{c}}^{\rm A}_{-1} = \bm{c}^{\rm B}_{-1} = \bm{0}$ and ${\bm{s}}^{\rm A}_{-1}=\bm{s}^{\rm B}_{-1} = \bm{0}$  following the convention in LSTM networks. 

The idea is to train the LSTM to capture the useful information in the sequence of historical observations into its cell state and hidden state. We then use the hidden state vectors ${\bm s}^{\rm A}_{\ell}$ and $\bm{s}^{\rm B}_{\ell}$ to design the corresponding sensing vectors. This is achieved by using fully connected DNNs. As shown in Fig.~\ref{fig:lstm_framework}, at agent B, the hidden state vector ${\bm s}^{\rm B}_{\ell}$ is taken as input to two different fully connected DNNs to output the next transmit sensing beamforming vector ${\bm w}^{\rm B}_{{\rm t},\ell}$ and receive sensing beamforming vector $\bm w^{\rm B}_{{\rm r},\ell+1}$ as follows:
\begin{subequations}
    \begin{align}
        &{\bm w}^{\rm B}_{{\rm t},\ell} = f_{\rm t, DNN}^{\rm B} ({\bm s}^{\rm B}_{\ell}),\\
        &\bm w^{\rm B}_{{\rm r},\ell+1} = f_{\rm r, DNN}^{\rm B} ({\bm s}^{\rm B}_{\ell}).
    \end{align}
\end{subequations}
Analogously, the next transmit and receive sensing beamformers at agent A, i.e., $\bm w^{\rm A}_{{\rm t}, \ell+1}$ and ${\bm w}^{\rm A}_{{\rm r},\ell+1}$, are designed by two different fully connected DNNs with the hidden state vector $\bm{s}^{\rm A}_\ell$ as input as shown in Fig.~\ref{fig:lstm_framework}.

The cell and hidden state vectors are closely related to the channel parameters' posterior probability, which is a sufficient statistic of received pilots. In \cite{8792366,9448070}, which focus on the AoA estimation problem for a single path channel, posterior probability is used to summarize the information from received pilots. However, computing the posterior probability for a multipath channel is challenging.
The main advantage of our proposed approach is that the LSTM cell can automatically and efficiently learn to update the cell/hidden state vectors from the sequence of observations. 

Fig.~\ref{fig:lstm_framework} illustrates the proposed active sensing unit in the $\ell\text{-th}$ pilot transmission round. To train the neural networks in the active sensing unit, we concatenate $L$ active sensing units together to form a very deep neural network, corresponding to the $L$ rounds of pilot transmission in Fig.~\ref{fig:pilot_transmission_pingpong1}. The overall deep active sensing architecture is shown in Fig.~\ref{fig:overall_lstm}. The neural network parameters across the different pilot transmission rounds can be tied together to reduce the training complexity. After $L$ rounds of pilot transmission, the cell state vector $\{\bm c^{\rm A}_{L-1}, \bm{c}^{\rm B}_{L-1}\}$ of the LSTMs at both sides are respectively mapped to the final beamforming vectors $\{\bm w_{{\rm t}}, \bm w_{\rm r}\}$  for the data transmission phase. This is done by employing another two DNNs in the final stage as follows:
\begin{subequations}\label{eq:final_dnn_mmwave}
    \begin{align}
        &\bm{w}_{\rm t} = g_{{\rm t, DNN}} (\bm{c}^{\rm A}_{L-1}),\\
        &\bm{w}_{\rm r} = g_{{\rm r, DNN}} ({\bm c}^{\rm B}_{L-1}).
    \end{align}
\end{subequations}
The overall neural network can be trained end-to-end in order to maximize the utility function $\mathbb{E}\left[|\bm w_{\rm r}^{\sf H}\bm G^{\sf H}\bm w_{\rm t}|^2\right]$ by employing stochastic gradient descent (SGD). In this way, the active sensing strategies together with the final DNNs at both sides are jointly optimized. Once trained and deployed, no feedback is needed between the Tx and the Rx.

\begin{figure*}[t]
    \begin{minipage}[b]{0.48\linewidth}
        \centering
        \includegraphics[width=8cm]{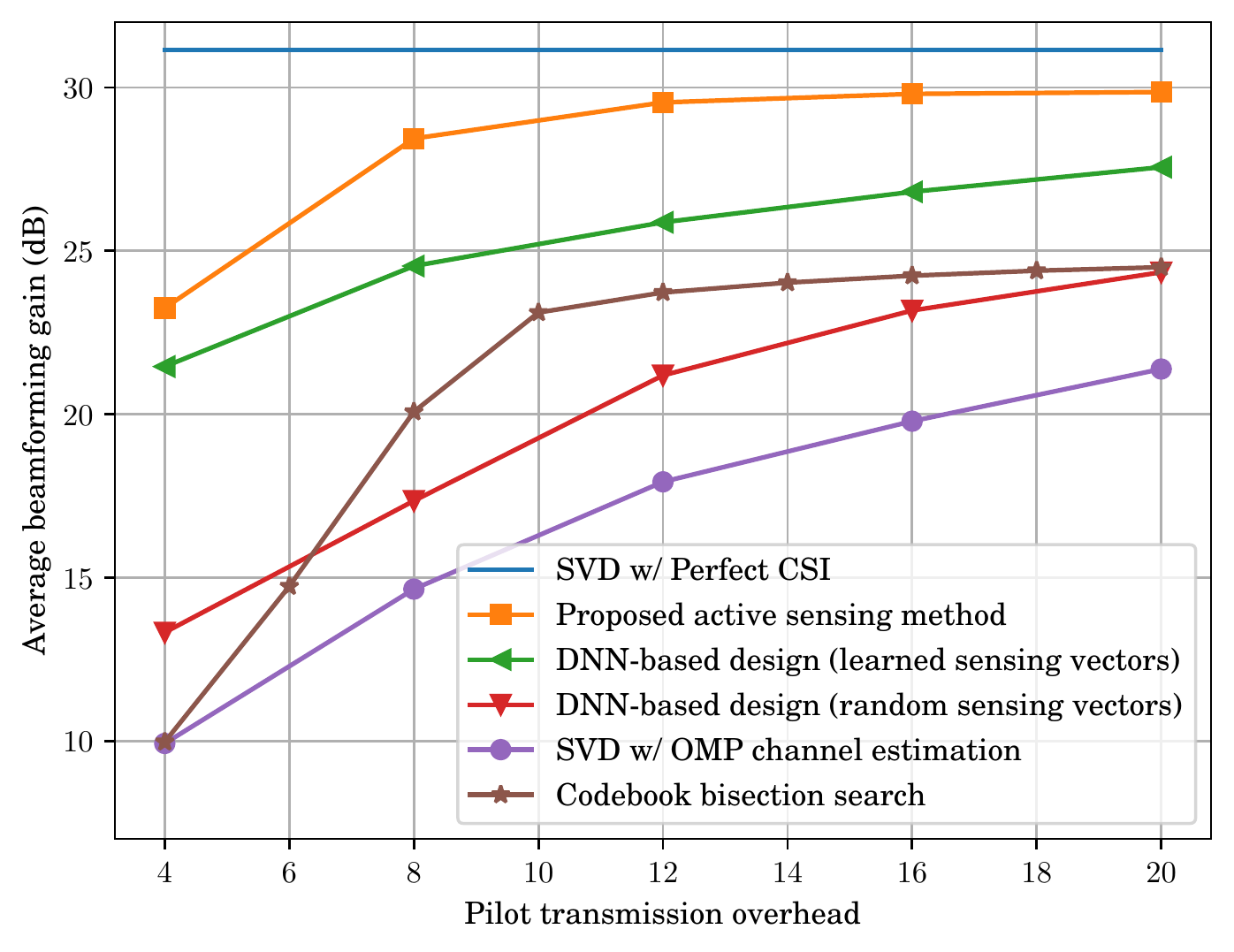}
        \caption{Average beamforming gain vs. pilot training overhead ($2L$) for a system with $M_{\rm t}=64$, $M_{\rm r}=32$, and $P_1/\sigma^2 = P_2/\sigma^2=0$ dB.}\label{fig:bf_gain_hybrid}
    \end{minipage}\hspace{0.5cm}
    \begin{minipage}[b]{0.48\linewidth}
        \centering
        \includegraphics[width=8cm]{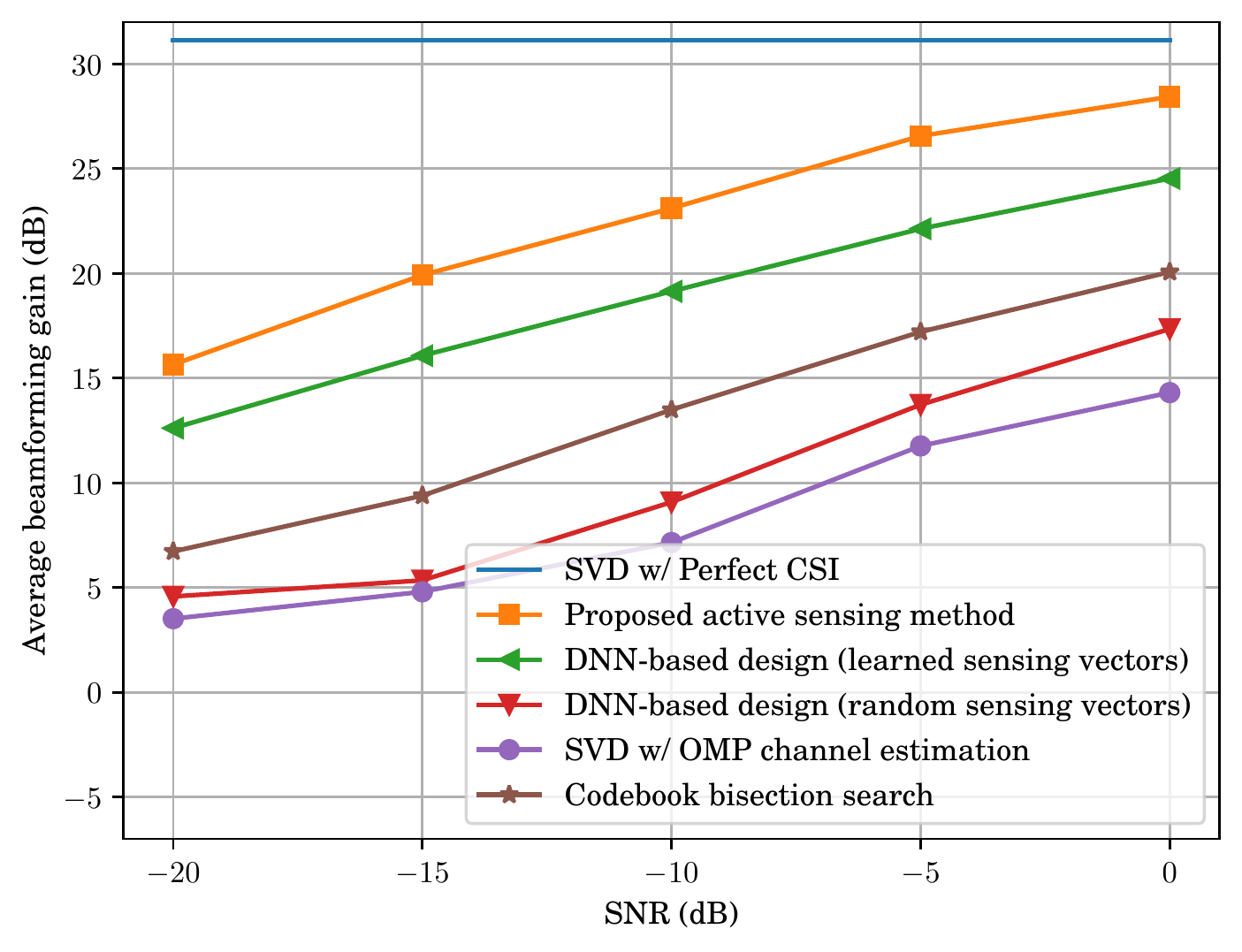}
        \caption{Average beamforming gain vs. SNR, where $\text{SNR} \triangleq P_1/\sigma^2=P_2/\sigma^2$ for a system with $M_{\rm t}=64, ~M_{\rm r}=32$, and $L = 4$.}\label{fig:bf_gain_hybrid_SNR}
    \end{minipage}
\end{figure*}

\section{Performance Evaluation for Two-Sided Beam Alignment}
\label{sec:sim_hybrid}

In this section, we evaluate the performance of the proposed active sensing method for the two-sided beam alignment problem and interpret the learned beamforming patterns.

In particular, we consider a system with $M_{\rm t}=64$ antennas at the Tx and $M_{\rm r}=32$ antennas at the Rx. The number of paths between the Tx and Rx is set to $L_p=3$. For each channel realization, the AoAs/AoDs are uniformly generated from $[-60^\circ, 60^\circ]$, and the complex fading coefficients are randomly taken from the distribution $\mathcal{CN}(0,1)$. In the pilot training phase, the raw SNRs in both directions, i.e., $P_1/\sigma^2$ and $P_2/\sigma^2$, are set to be $0$dB unless otherwise stated.

\subsection{Implementation Details}\label{sec:impl_hybrid}
For the proposed deep active sensing unit, the dimensions of hidden states and cell states are both set to be $512$ for the Tx side and both set to be $256$ for the Rx side. The DNNs $f_{\rm t, DNN}^{\rm A}(\cdot)$ and $f_{\rm r, DNN}^{\rm A}(\cdot)$ are both of size $[512,512,2M_{\rm t}]$, and the size of DNN $g_{\rm t, DNN}(\cdot)$ is  $[1024,1024,2M_{\rm t}]$; the DNNs $f_{\rm t,DNN}^{\rm B}(\cdot)$ and $f_{\rm r, DNN}^{\rm B}(\cdot)$ are both of size $[512,512,2M_{\rm r}]$, and the size of DNN $g_{\rm r, DNN}(\cdot)$ is $[1024,1024,2M_{\rm r}]$. We adopt the rectified linear unit (ReLU) activation function in all the dense layers except the last layer that outputs the beamforming vector. Since most current deep learning software packages only support real-valued operations, the active sensing unit outputs a real-valued vector that contains the real and imaginary parts of the beamforming vector. To meet the unit $\ell_2$-norm constraint on the beamforming vectors, the activation function\footnote{For the phase-only beamformers, the activation function can be chosen as $\varrho(\cdot) = {\cdot}/{|\cdot|}$, which applies element-wise modulus normalization to the complex beamforming vector.} 
for the last dense layer is chosen to be $\varrho(\cdot) = {\cdot}/{\|\cdot \|_2}$.  A batch normalization layer is added between two dense layers in the fully connected neural networks to accelerate the training process \cite{ioffe2015batch}. The initial sensing vectors $\bm w^{\rm A}_{{\rm t},0}$, $\bm w^{\rm B}_{{\rm r},0}$, and $\bm w^{\rm A}_{{\rm r},0}$ as shown in Fig.~\ref{fig:pilot_transmission_pingpong1} are set as trainable parameters in the implementation, so that the optimizer can update them in the training stage. This means that they are learned from the distribution of the training data.  The parameters of the dense layers in the active sensing unit are tied together across different sensing stages to reduce the training complexity.
The overall deep learning framework is implemented on Tensorflow \cite{abadi2016tensorflow} and trained by Adam optimizer \cite{kingma2014adam} with a learning rate progressively decreasing from $10^{-4}$ to $10^{-5}$. 
We generate (based on the channel model) as many training samples as needed to train the neural network, and stop training when the performance on a validation set of $10^4$ samples is no longer increasing over several epochs. The testing dataset consists of $10^4$ i.i.d. channel realizations.

\subsection{Benchmarks}
We compare the performance of the proposed method with the following benchmarks. For most of these benchmarks (except the bisection search scheme), the beamformers are all designed at one side, so error-free feedback is needed to communicate the designed beamformer to the other side. In the pilot training phase, a total of $2L$ pilot symbols are sent as compared to the proposed methods with $L$ rounds of ping-pong pilot transmission.

\emph{Compressive sensing with random vectors \cite{6847111}:} This method adopts the compressive sensing method in which the orthogonal matching pursuit (OMP) algorithm is used \cite{4385788} to estimate the channel matrix $\bm G$. The sensing vectors are generated randomly. Given the estimated channel, the beamformers for data transmission are given by the singular value decomposition (SVD) method as in \eqref{eq:optimal_bf}. 

\emph{DNN-based design with random sensing vectors \cite{attiah2021deep,9427148}:} In this approach,  we train two fully connected neural networks to directly map the received pilots to the optimized beamformers of the Tx and the Rx, respectively, in a centralized fashion. The DNNs for the Tx and Rx are of size $[1024,1024,2M_{\rm t}]$ and $[1024,1024,2M_{\rm r}]$, respectively. The sensing vectors are generated randomly and fixed for all the channel realizations.

\emph{DNN-based design with learned sensing vectors:} In this approach,  we use the same neural network architectures as in the DNN-based design with random sensing vectors, but the sensing vectors are learned from the training data. 
In other words, the sensing vectors are designed according to the channel statistics, rather than as functions of the received pilots, which contain information about the instantaneous CSI, as in the proposed active sensing approach. 
To this end, we make the sensing beamformers as trainable layers in implementing the neural network so that the SGD algorithm can update these sensing beamformers in the training phase. After training, these sensing beamformers are fixed for all the channel realizations.

\emph{Codebook bisection search \cite{8644444}:} The scheme uses a predefined hierarchical codebook to choose the sensing beamformers at both sides. Specifically, upon receiving a pilot signal, the transmitter/receiver sorts the received signal power of all the pilots received so far, and then splits the beam into narrower beams in the direction with the strongest received power. 

\subsection{Simulation Results}
\label{subsect:sim_hybrid}
The simulation results are illustrated in Figs.~\ref{fig:bf_gain_hybrid} and \ref{fig:bf_gain_hybrid_SNR}. It can be seen that the data-driven approaches, which bypass channel estimation, significantly outperform the channel estimation based approach. The DNN-based approach with learned sensing vectors can achieve much better performance as compared to the DNN-based method with random sensing vectors. This implies that designing the sensing vectors is important for improving the quality of the received pilots. 
Further, the proposed active sensing approach, which utilizes pilot observations that are functions of instantaneous CSI, achieves better performance than the DNN-based approach that learns the sensing vectors based on channel statistics. 
Specifically, at raw SNR of 0dB, the proposed approach using $8$ pilot transmission overhead can outperform all the other benchmarks with $20$ pilot transmissions. This indicates that the proposed deep active sensing framework can efficiently design the sensing vectors from the observations, thereby reducing the pilot training overhead. We observe that the codebook bisection search benchmark can outperform some non-adaptive sensing approaches, but its performance is significantly lower than the proposed active sensing approach.

We remark that the proposed scheme does not require feedback of the designed beamformers from one side to the other either. However, feedback is necessary for most benchmarks, which would require implementation of an extra side channel between the transmitter and the receiver. If feedback is allowed in the sensing stage, we can readily modify the proposed two-sided scheme to incorporate the feedback information to gain further performance improvements.

In Fig.~\ref{fig:bf_gain_hybrid_SNR}, we further show the average beamforming gain against the SNR in the pilot training phase. The SNR is set to be the same in both pilot transmission directions, i.e., $\text{SNR}\triangleq P_1/\sigma^2=P_2/\sigma^2$. From Fig.~\ref{fig:bf_gain_hybrid_SNR}, it can be seen that the proposed active sensing method consistently outperforms the other benchmarks. Even when the SNR is $-20$dB, the proposed active sensing approach still achieves better performance than the conventional channel estimation based approach at $0$dB. It should be noted that for the sake of simplicity, the SNR is set to be the same for both sides here. However, the proposed approach still holds advantages if the SNR values for the two sides are different.

\begin{figure}[t]
    \centering
    \includegraphics[width=8cm]{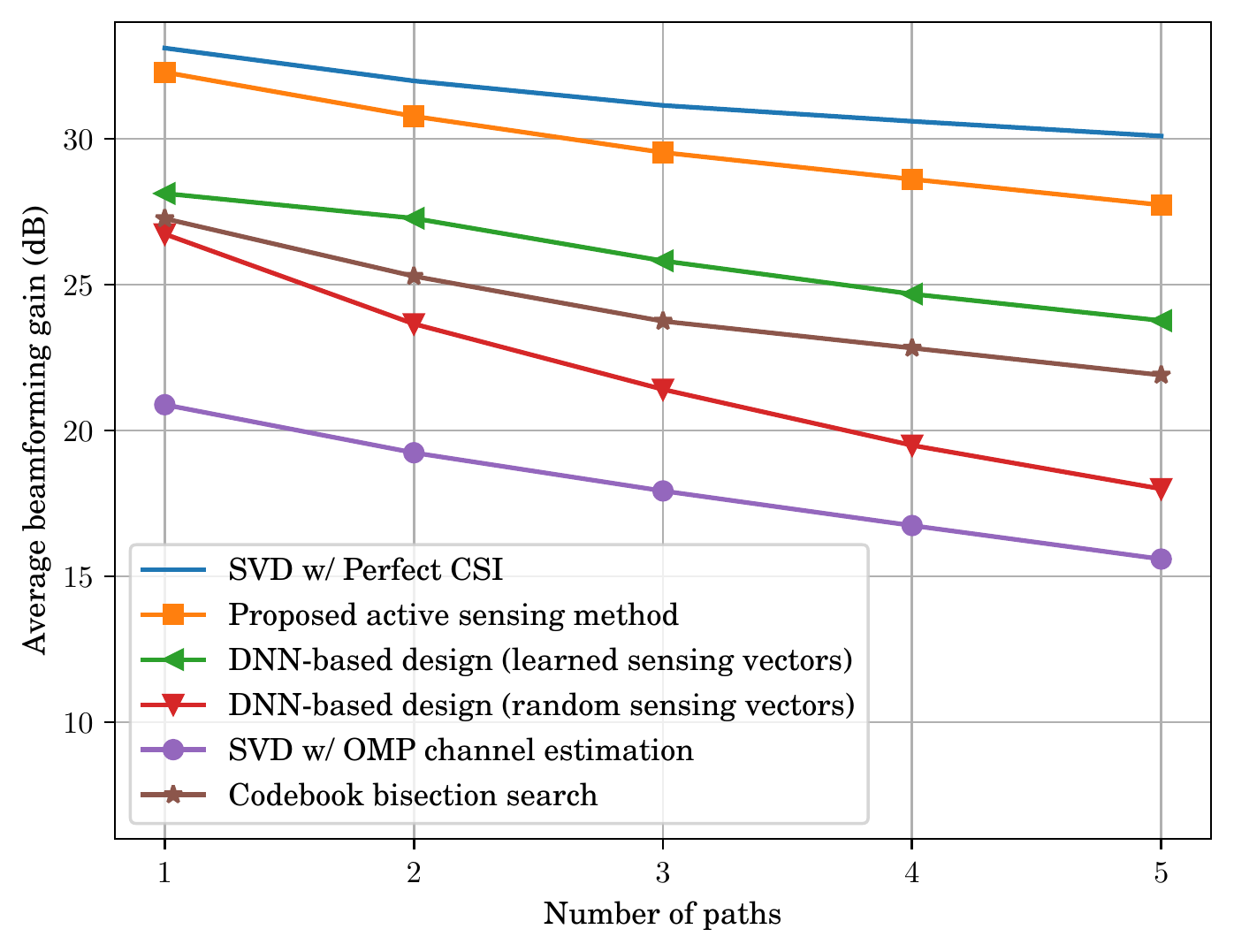}
    \caption{Generalization performance to different number of paths. All the neural networks are trained when the number of path is $3$.}
    \label{fig:generalization}
\end{figure}

In Fig.~\ref{fig:generalization}, we investigate the generalization ability of the proposed methods when the number of path varies. All the neural network based approaches are trained using channel realizations with three paths. The results show that the proposed active sensing method consistently outperforms the other benchmarks even as the number of paths varies, indicating that the neural network generalizes well. To further improve the generalization performance, we can incorporate different channel scenarios in the training data. For example, the training data can contain channels with differing numbers of paths or a range of different SNRs. In this way, the DNN can be made to adapt to the different system parameter settings more easily.

\subsection{Interpretation of Solutions from Active Sensing}
\begin{figure*}
    \centering
    \includegraphics[width=18cm]{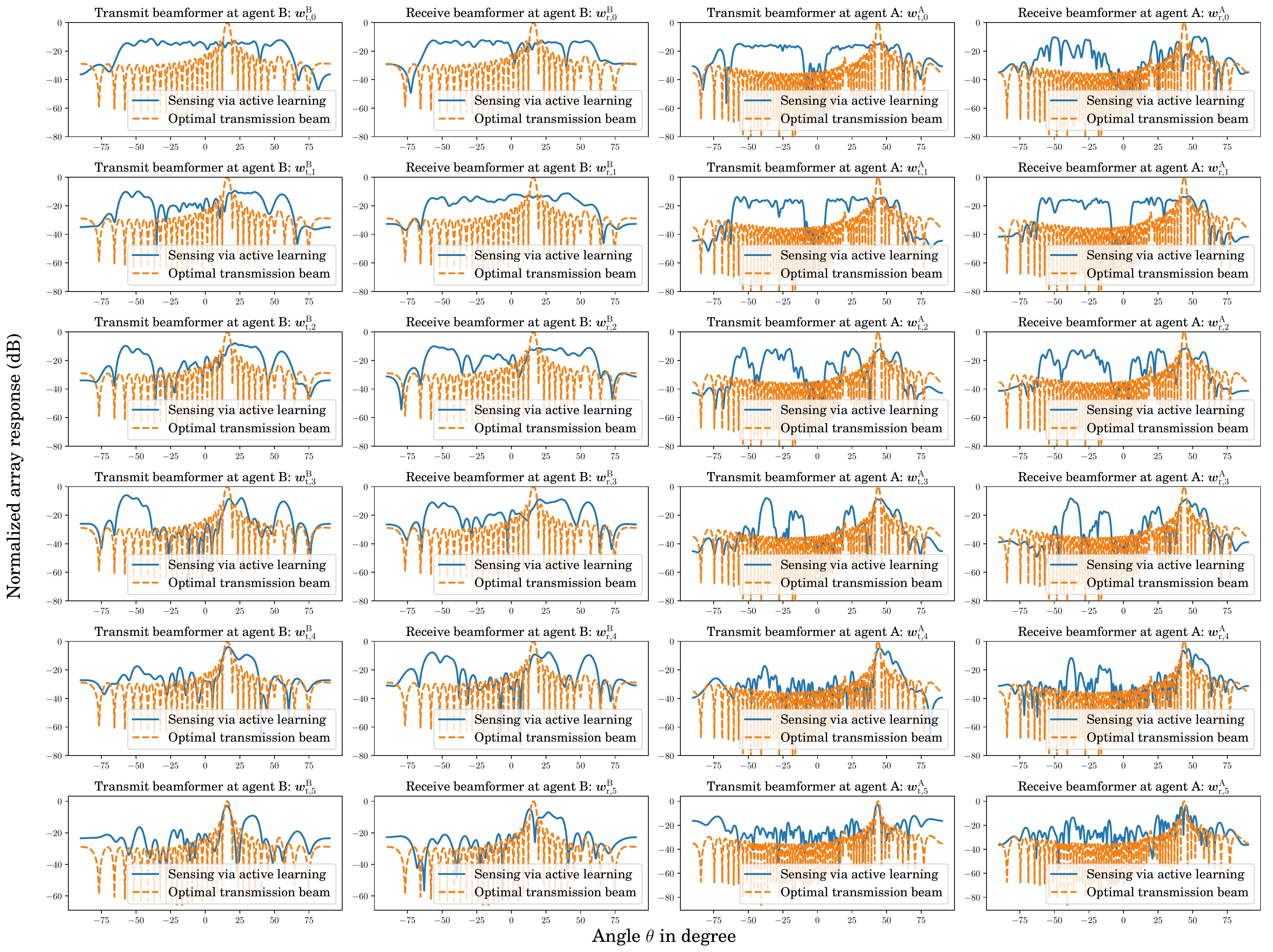}
    \caption{Learned sensing beamforming patterns for a specific channel realization with $L=6$, $M_{\rm t}=64$, $M_{\rm r}=32$ and $P_1/\sigma^2=P_2/\sigma^2=0$~dB.}\label{fig:bf_pattern6_9}
    \vspace{0.3cm}
    \subfigure[Example of beamformers matching the strongest singular-vector direction]{\includegraphics[width=8.8cm]{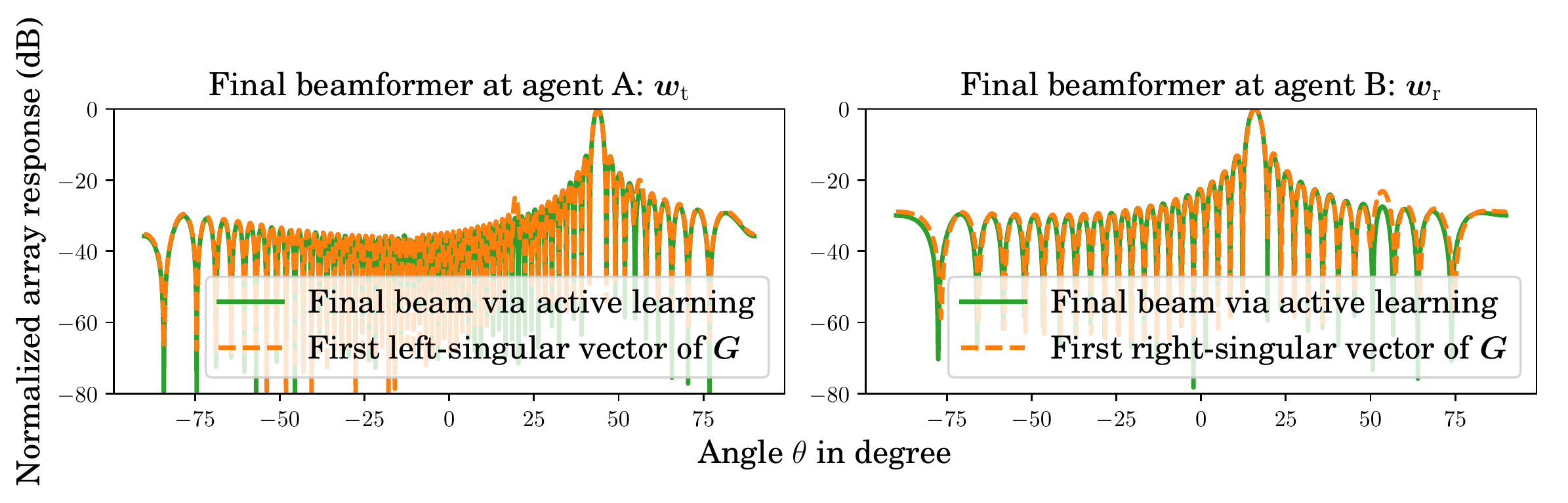}\label{fig:bf_pattern_data6_9}}\hspace{0.2cm}
    \subfigure[Example of beamformers matching the second singular-vector direction]{\includegraphics[width=8.8cm]{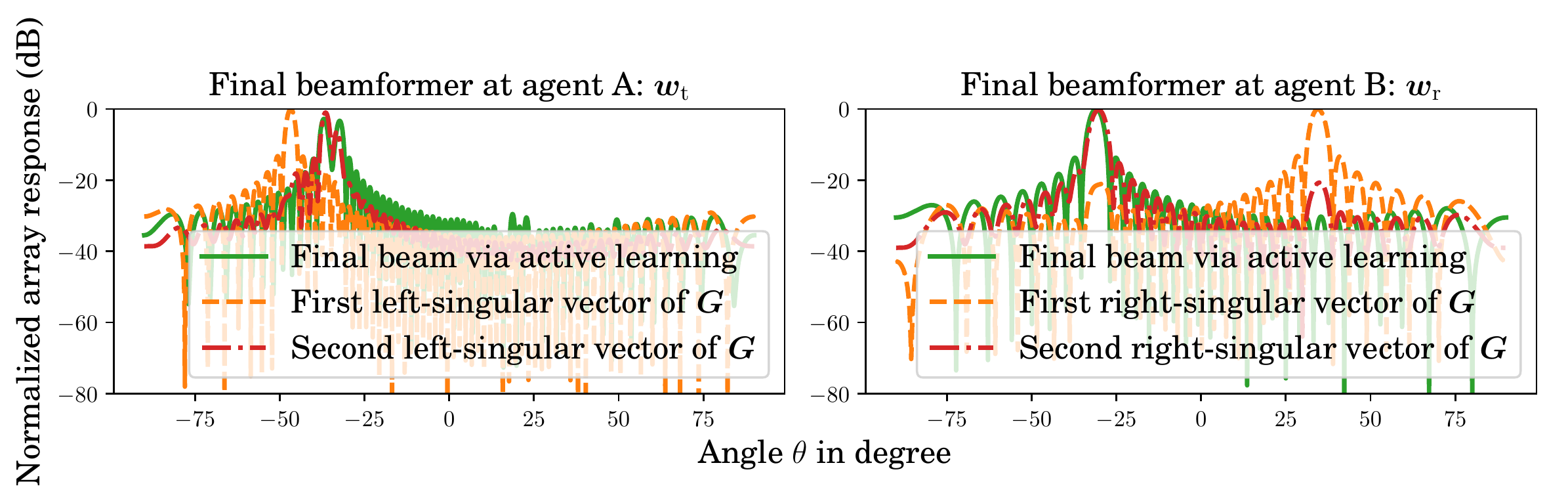}\label{fig:bf_pattern_data6_20}}
    \caption{Learned final data transmission beamforming patterns for two channel realizations with $L=6$, $M_{\rm t}=64$, $M_{\rm r}=32$ and $P_1/\sigma^2=P_2/\sigma^2=0$~dB.}
    \vspace{0.3cm}
    \includegraphics[width=18.2cm]{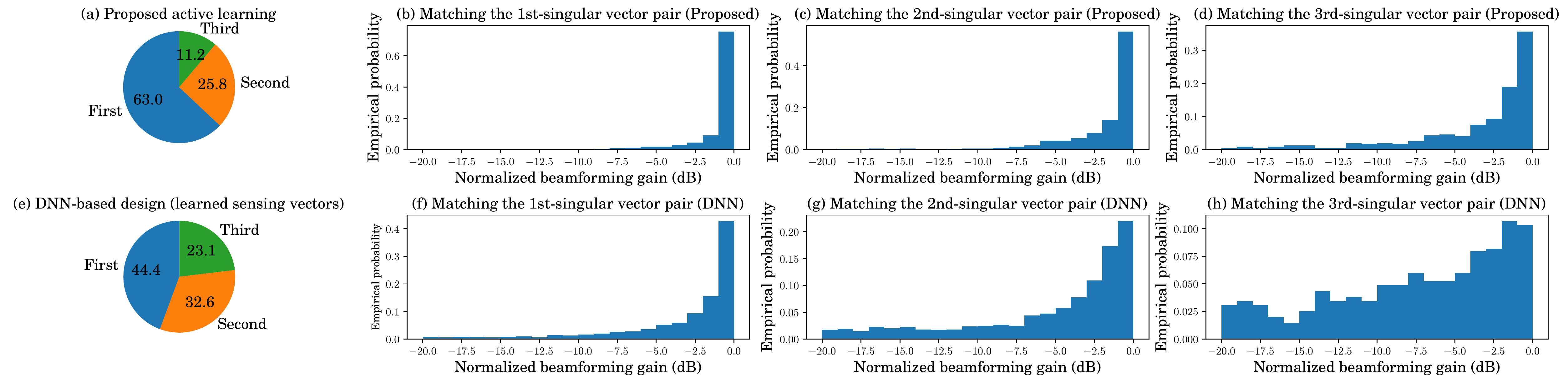}
    \caption{Proportion of mmWave channel realizations for which the proposed active sensing method and the DNN-based method with sensing vectors learned from the channel statistics, respectively, match the first, second and third singular vectors, and the histograms of the beamforming gains in each case. The system setting is $L=6$, $M_{\rm t}=64$, $M_{\rm r}=32$ and $P_1/\sigma^2=P_2/\sigma^2=0$~dB.}\label{fig:bf_gain_pie6}
\end{figure*}

We have shown that the proposed active sensing method can outperform the other
benchmarks for the two-sided beam alignment problem. It is important to
interpret the solutions learned via the LSTM based neural networks and to
understand where the gains come from. To this end, we examine
the beamforming patterns of the designed sensing and beamforming
vectors in this section. For a sensing/beamforming vector $\bm w\in\mathbb{C}^{M}$, the
normalized array response  under such configuration is defined as
\begin{align}\label{eq:bf_gain_effective}
    f_{\rm beam}(\bm w,\theta) = |\bm w^{\sf H}\bm a(\theta)|^2,~~\forall\theta\in[-\pi/2,\pi/2],
\end{align} 
where $\bm a(\theta)=\frac{1}{\sqrt{M}} [1,\cdots,e^{j\pi(M-1)\sin(\theta)}]^\top$ is the normalized steering vector. 

In Fig.~\ref{fig:bf_pattern6_9}, we plot the beamforming pattern of the sensing vectors $\bm w^{\rm B}_{{\rm r},\ell}$, ${\bm w}^{\rm B}_{{\rm t},\ell}$, $\bm w^{\rm A}_{{\rm t},\ell}$ and ${\bm w}^{\rm A}_{{\rm r},\ell}$ designed via the proposed active sensing framework for $\ell=0,\cdots,5$ for a randomly generated channel realization. The active sensing neural network is trained for $L=6$ ping-pong transmission rounds. As can be seen from Fig.~\ref{fig:bf_pattern6_9}, the sensing vectors designed by the active sensing unit have relatively uniform array response in the first two rounds, i.e., $\ell=0,1$. As the number of observations increases, the active sensing units at both sides gradually learn to narrow down the search directions and to focus more energy in the optimal direction, but they also try to explore other directions. 
This can be interpreted as a behavior of systematic exploration of channel landscape, learned by the proposed active sensing method for finding the optimal beamformers for the eventual data transmission. 

In Fig.~\ref{fig:bf_pattern_data6_9}, we plot the beamforming pattern of the data transmission beamforming vectors $\{\bm w_{\rm t}, \bm w_{\rm r} \}$, which are learned via the DNNs after the final sensing stage as in \eqref{eq:final_dnn_mmwave} for two different channel realizations. 
In Fig.~\ref{fig:bf_pattern_data6_9}, it can be seen that the final data transmission beamformers perfectly match the optimal beamforming vectors computed from \eqref{eq:optimal_bf} assuming perfect CSI, (i.e., the singular vectors corresponding to the largest singular value). 
This means that the proposed active sensing framework indeed learns an intelligent strategy to design the sensing vectors to collect the received pilots so that the optimal final data transmission beamformers can be found within a few pilot transmission rounds.

In Fig.~\ref{fig:bf_pattern_data6_20}, we show that for a different channel realization 
the beamforming pattern designed by the same neural network happens 
to match the singular vectors corresponding to the second-largest singular value
of the channel matrix $\bm G$. In this case, the final learned transmission
beamformer is suboptimal. This phenomenon accounts for the gap between the proposed
active sensing method and the perfect-CSI benchmark seen in
Fig.~\ref{fig:bf_gain_hybrid}.  This gap is approximately $1.58$dB when 
the number of pilot transmission overhead is $12$ (or $L=6$). 

To see how often the learned data transmission beamformers match the singular
vectors corresponding to the largest singular value versus the second or the third
largest singular values, we calculate the beamforming gains in the different
effective directions for 10,000 random realizations of the channel and summarize their
statistics in Fig.~\ref{fig:bf_gain_pie6}. More specifically, we assume a channel model
with three paths, i.e., $\bm G = \sum_{i=1}^3\sigma_i \bm u_i\bm v_i^{\sf H}$, where 
$\sigma_i$ is the $i$-th largest singular value and the associated unit-norm singular vectors 
are $\{\bm u_i,\bm v_i\}$, i.e., $\|\bm u_i\|_2=1$ and $\|\bm v_i\|_2=1$. 
We define the beamforming gain in the $i$-th effective direction as 
\begin{align}
    f_{\rm BA}^i(\bm w_{\rm t}, \bm w_{\rm r}) = |\bm w_{\rm t}^{\sf H} \bm u_i\bm v_i^{\sf H}\bm w_{\rm r}|^2.
\end{align}
Note that the maximum of $f_{\rm BA}^i(\bm w_t, \bm w_r)$ is $0$dB, which is achieved when the beamformers $\{\bm w_{\rm t}, \bm w_{\rm r}\}$ match the effective directions $\{\bm u_i,\bm v_i\}$ perfectly, i.e., $\bm w_{\rm t} = \bm u_i$ and $\bm w_{\rm r}=\bm v_i$.

Given the final beamformers $\{\bm w_{\rm t}, \bm w_{\rm r}\}$ learned by the proposed active sensing method, we calculate the beamforming gain in each of the effective directions to see in which direction the beamformer pairs have the largest beamforming gain. 
As can be seen from Fig.~\ref{fig:bf_gain_pie6}(a), among these three effective directions, the beamforming gain is maximized in the first, second, and third effective directions $63.0\%$, $25.8\%$, $11.2\%$ of the time, respectively, for the proposed active sensing approach. 
In contrast, as shown in Fig.~\ref{fig:bf_gain_pie6}(e), if the sensing vectors are designed from channel statistics as in DNN-based design baseline, the probability of matching to the largest singular value is significantly reduced. 

Figs.~\ref{fig:bf_gain_pie6}(b)(c)(d) further show the histograms of the actual beamforming gains within each of the three cases. Here, 0dB indicates that the learned sensing vector matches the respective effective direction perfectly. As seen in the figure, the learned sensing beamformers match the effective direction most of the time for the proposed active sensing approach. 
In comparison, for the benchmark of the DNN-based design with sensing vectors learned from channel statistics, Figs.~\ref{fig:bf_gain_pie6}(f)(g)(h) show that mismatches occur more often. 

We remark that this paper focuses on reducing the communication cost of pilot transmission. It is important to note that the computation cost of neural network processing should also be considered. The neural networks are trained offline. The proposed active sensing framework has a longer training time than the DNN-based design. We compare the inference time complexity in Table \ref{table}, which indicates that the proposed method has comparable time complexity as compared to other benchmarks. At the inference stage, its computational
complexity also depends on hardware implementation and can be reduced through further optimization of the neural network. In the proposed framework, the computational complexity is dominated by the matrix-vector product operation, which can be efficiently implemented in parallel using hardwares such as GPUs and FPGAs. Additionally, several compression techniques \cite{9134426}, such as DNN sparsification \cite{8682464} can be applied to reduce the computational complexity of the current architecture without affecting performance. The proposed framework is expected to compare favorably with the computational complexity of benchmark algorithms in the implementation stage.

\begin{table}[t]
    \caption{Inference time complexity. Here, $M = \max(M_{\rm t},M_{\rm r})$, $h$ is the hidden layer size of neural networks.\label{table}
    }
    \centering
    \begin{tabular}{ |c|c|c|c| } 
     \hline
     Proposed & DNN-based design & OMP+SVD & Bisection  \\ 
     \hline
     ${O}(LMh+Lh^2)$  & ${O}(Lh+Mh+h^2)$ & ${O}(LM_{\rm t}M_{\rm r})$& ${O}(\log M)$\\
     \hline
    \end{tabular}
\end{table}

\section{Two-sided Beam and Reflection Alignment}
\label{sec:sys_ris}
\begin{figure}[t]
    \centering
    \includegraphics[width=6cm]{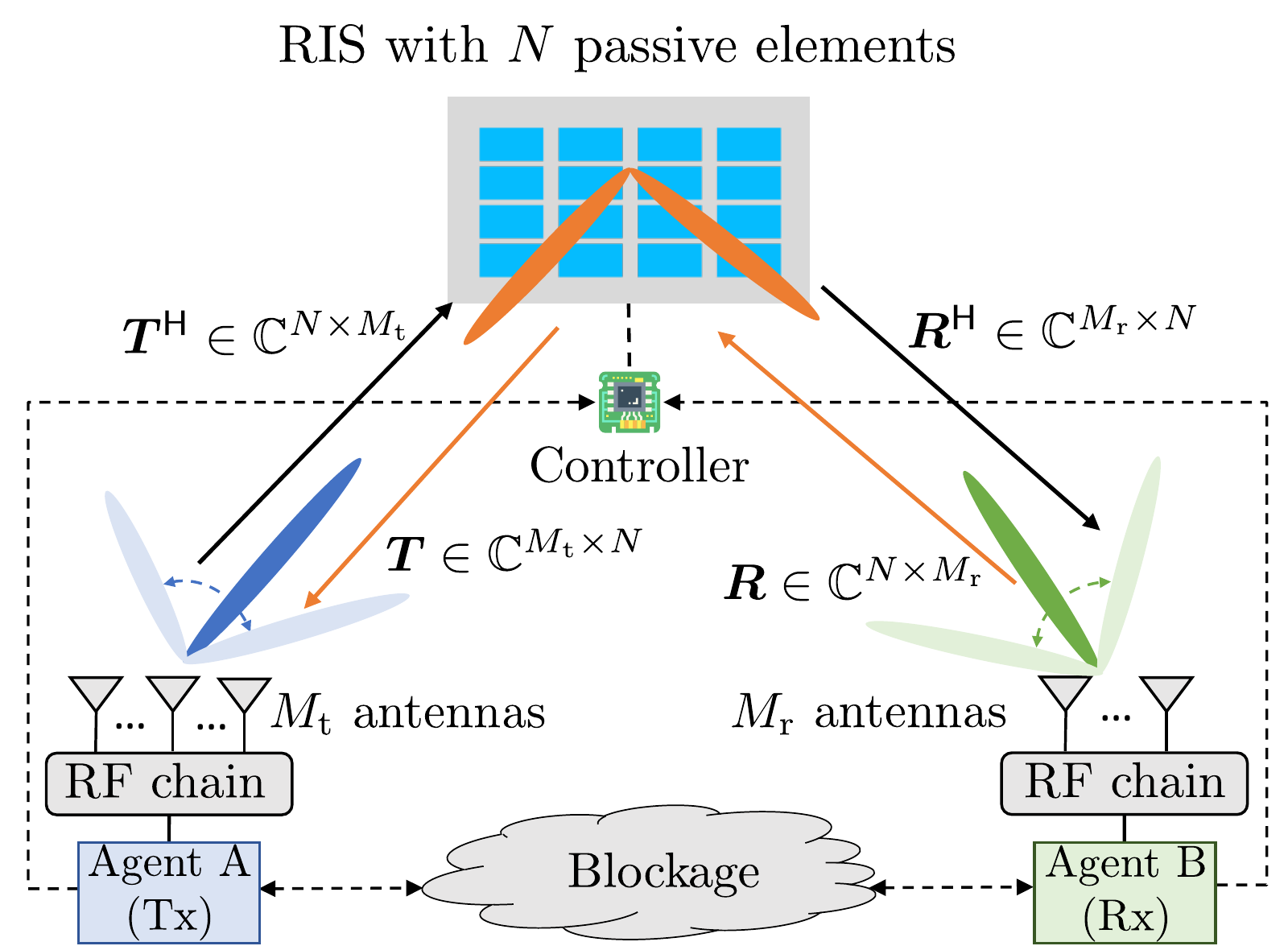}
    \caption{Two-sided beam and reflection alignment problem in the RIS system.}\label{fig:system_model}
\end{figure}

The proposed RNN-based active sensing framework can be extended to the
scenario where an RIS is placed between the Tx and the Rx to tackle the blockage
issue in mmWave communications. In particular, we show that the proposed active
sensing strategy can be used to adaptively configure the RIS reflection coefficients along with the transmit and receive beamformers
for both sensing and communication.

\subsection{System Model}
Consider an RIS-assisted mmWave MIMO communication system including a Tx (agent A) with $M_{\rm t}$ antennas as a ULA, an Rx (agent B) with $M_{\rm r}$ antennas as a ULA, and a uniform rectangular array RIS with $N$ passive elements. The Tx and the Rx each have a single RF chain. The direct channel between the Tx and the Rx is blocked due to an unfavorable propagation environment. 
The RIS is deployed between the Tx and the Rx to enable a signal reflection path, as shown in Fig.~\ref{fig:system_model}.  
The reflection coefficients of the RIS are denoted by $\bm v=[e^{j\omega_1},\cdots,e^{j\omega_N}]^\top\in\mathbb{C}^N$, where $\omega_i$ is the phase shift of the $i$-th element. As before, let $\bm w_{\rm t}\in\mathbb{C}^{M_{\rm t}}$ and $\bm w_{\rm r}\in\mathbb{C}^{M_{\rm r}}$ denote the beamforming vectors at the Tx and the Rx side, respectively, 
with $\|\bm w_{\rm t}\|_2=\|\bm w_{\rm r}\|_2=1$.  To establish a reliable link between the Tx and the Rx, the beamforming vectors $\{\bm w_{\rm t}, \bm w_{\rm r}\}$ and the reflection coefficients $\bm v$ should be jointly optimized according to the CSI, so that the SNR of the communication link is maximized. We refer to this as the beam and reflection alignment problem.

As shown in Fig.~\ref{fig:system_model}, let the matrix $\bm T\in\mathbb{C}^{M_{\rm t} \times N}$ denote the channel from the RIS to the Tx, and $\bm R\in\mathbb{C}^{ N\times M_{\rm r} }$ denote the channel from the Rx to the RIS.  As before, the channels are modeled by a sparse multipath channel model, namely: 
\begin{subequations}
    \begin{align}
        &\bm T = \sum_{i=1}^{L_{p_{\rm t}}}\alpha_{\rm t}^i \bm a_{\rm t}(\phi_{\rm t}^i, \theta_{\rm t}^i)\bm a_{\rm v}^{\sf H}(\phi_{\rm v}^i, \theta_{\rm v}^i),\\
        &\bm R= \sum_{i=1}^{L_{p_{\rm r}}}\alpha_{\rm r}^i \bm a_{\rm v}(\phi_{\rm v}^i, \theta_{\rm v}^i)\bm a_{\rm r}^{\sf H}(\phi_{\rm r}^i, \theta_{\rm r}^i),
    \end{align}
\end{subequations}
where $L_{p_{\rm t}}$ and $L_{p_{\rm r}}$ denote the number of paths between the Tx/Rx and the RIS, respectively, and $\{\alpha^i_{\rm t}$, $\alpha^i_{\rm r}\}$ denote the corresponding complex fading coefficients of the $i$-th path, $\{\phi_{\rm t}^i, \theta_{\rm t}^i, \phi_{\rm v}^i, \theta_{\rm v}^i, \phi_{\rm r}^i, \theta_{\rm r}^i \}$ denote the corresponding azimuth and elevation angle-of-arrival/angle-of-departure (AoA/AoD) of the $i$-th path, and $\bm a_{\rm t}(\cdot),\bm a_{\rm r}(\cdot),\bm a_{\rm v}(\cdot)$ are the steering vectors of the Tx, the Rx and the RIS, respectively. 

We assume particular orientations of the Tx and the Rx in relation to the RIS
in a 3D geometrical model of their respective locations, so that the steering vectors
of the Tx, the Rx and the RIS can be written as functions of their respective
azimuth and elevation angles as follows 
\cite{9427148}:
\begin{subequations}
    \begin{align}
        &[\bm a_{\rm t}(\phi_{\rm t}^i, \theta_{\rm t}^i)]_m = e^{j\pi(m-1)\cos(\phi_{\rm t}^i)\cos(\theta_{\rm t}^i)}, \\
        &[\bm a_{\rm r}(\phi_{\rm r}^i, \theta_{\rm r}^i)]_m = e^{j\pi(m-1)\cos(\phi_{\rm r}^i)\cos(\theta_{\rm r}^i)}, \\
        &[\bm a_{\rm v}(\phi_{\rm v}^i, \theta_{\rm v}^i)]_m = e^{j\pi (i^\prime(m)\sin(\phi_{\rm v}^i)\cos(\theta_{\rm v}^i)+i^{\prime\prime}(m)\sin(\theta_{\rm v}^i))},
    \end{align}
\end{subequations}
where $i^\prime(m) = \mod(m-1,N_{\rm H})$ and $i^{\prime\prime}(m)=\lfloor (m-1)/N_{\rm H} \rfloor $ for $m=1,\cdots,N$, with $N_{\rm H}$ as the number of horizontal elements of the rectangular RIS. 
Here, all the antennas and the RIS elements are assumed to have half wavelength spacing. 

Then, the signal received at Rx can be written as
\begin{align}
    r = \bm w_{\rm r}^{\sf H}\bm R^{\sf H}\diag(\bm v^{\sf H})\bm T^{\sf H}\bm w_{\rm t}x+n,
\end{align}
The goal here is to maximize the beamforming gain, $|\bm w_{\rm r}^{\sf H}\bm R^{\sf H}\diag(\bm v^{\sf H})\bm T^{\sf H}\bm w_{\rm t}|^2$ by jointly optimizing the beamformers $\{ \bm w_{\rm r}, \bm w_{\rm t}\}$ and the reflection coefficients $\bm v$ according to the pilots received in the pilot training phase. 

\subsection{Ping-Pong Pilot Training Protocol}

\begin{figure}
\centering
    \includegraphics[width=8cm]{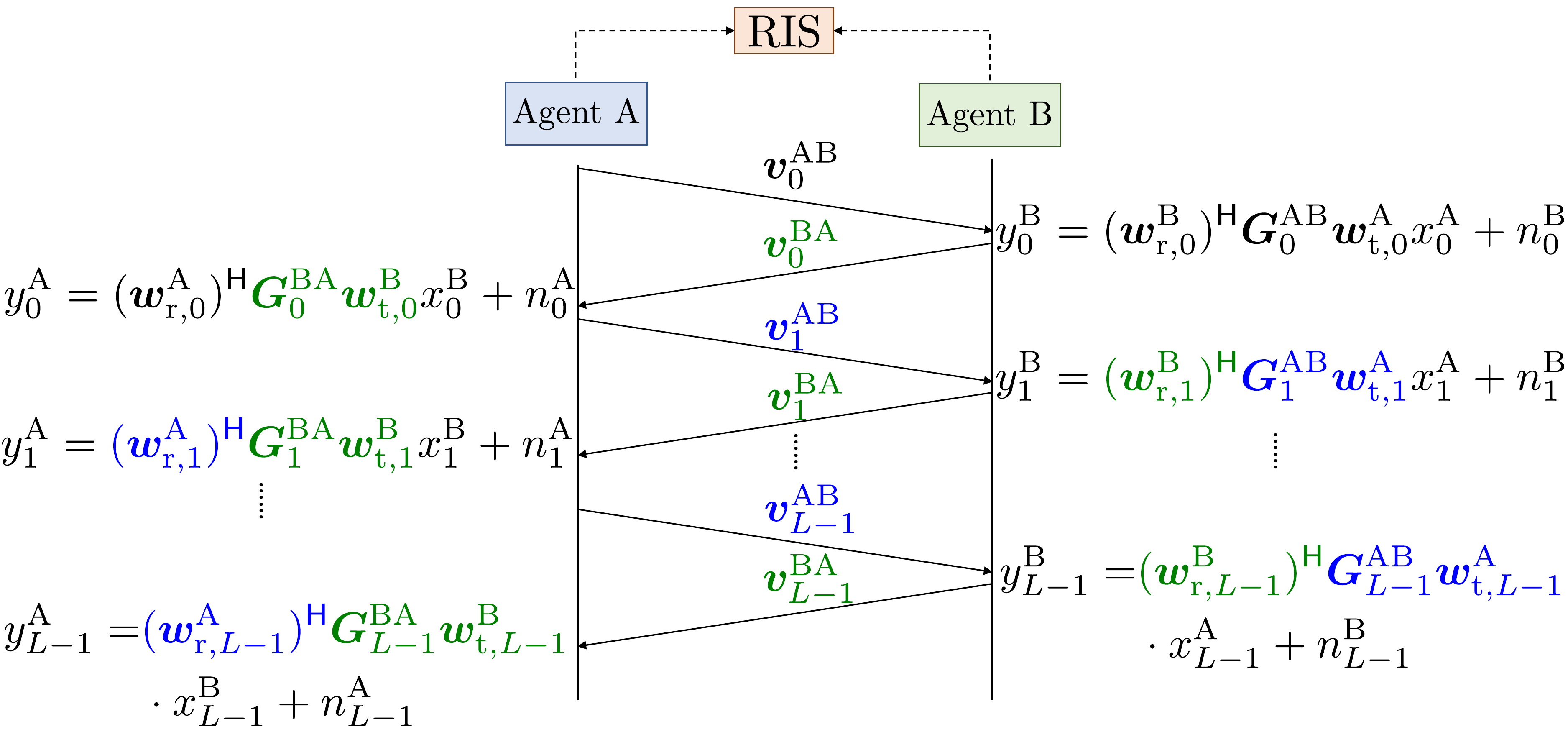}
    \caption{The proposed ping-pong pilot training protocol in the RIS-assisted mmWave system. The sensing beamformers actively designed at agent A and agent B are highlighted as blue (e.g., \textcolor{blue}{$\bm{w}^{\rm A}_{{\rm t},1},~\bm{w}^{\rm A}_{{\rm r},1}$}) and green (e.g., \textcolor{rx}{$\bm{w}^{\rm B}_{{\rm t},0},~\bm{w}^{\rm B}_{{\rm r},1}$}), respectively. The actively designed reflection coefficients are highlighted as green (e.g., \textcolor{rx}{$\bm v_0^{\rm BA}$}) if the latest observation comes from agent B, and as blue (e.g., \textcolor{blue}{$\bm v_1^{\rm AB}$}) if from agent A. The initial sensing vectors $\bm w^{\rm A}_{\rm t,0}$, $\bm w^{\rm A}_{\rm r,0}$, $\bm w^{\rm B}_{\rm r,0}$ and $\bm v_0^{\rm AB}$ are fixed and can be learned from the channel statistics.}
    \label{fig:pilot_transmission_pingpong_RIS}
\end{figure}

We consider a system setup in which an RIS controller collects information from both agent A and agent B to configure the RIS coefficients for both sensing and communications.\footnote{The proposed method can be easily modified to the scenario where only one side can control the RIS.} 
Similar to the pilot scheme proposed in Section~\ref{subect:ping-ping}, 
we propose a ping-pong pilot training scheme in which
agent A and agent B actively adjust their respective sensing beamformers locally, and the RIS controller adjusts the reflection coefficients based on information from both agents A and B for sensing. This is done over multiple rounds based on the observations collected so far as illustrated in Fig.~\ref{fig:pilot_transmission_pingpong_RIS}. 
At the end, the two agents produce their respective data transmission beamformers; the RIS controller adjusts the RIS reflection coefficients based on the inputs from agents A and B for data transmission.

More specifically, the received pilot symbol in the $\ell$-th round at agent B is given by 
\begin{align}
    y_\ell^{\rm B} = (\bm w^{\rm B}_{{\rm r},\ell})^{\sf H} \bm G^{\rm AB}_{\ell}\bm w^{\rm A}_{{\rm t},\ell} x^{\rm A}_\ell+n^{\rm B}_\ell,\quad\ell=0,\cdots,L-1,
\end{align}
where $\bm G^{\rm AB}_\ell \triangleq \bm R^{\sf H}\diag(\bm v^{\rm AB}_\ell)\bm T^{\sf H}$ is the cascaded channel from agent A to agent B with the configuration of RIS as $\bm v^{\rm AB}_\ell$, the vectors $\bm w^{\rm A}_{{\rm t},\ell}\in\mathbb{C}^{M_{\rm t}}$ and $\bm w^{\rm B}_{{\rm r},\ell}\in\mathbb{C}^{M_{\rm r}}$ are the sensing beamformers in the $\ell$-th round pilot transmission at agent A and agent B, respectively, and $n_\ell^{\rm B}\sim\mathcal{CN}(0,\sigma^2)$ is the additive Gaussian noise. 
Note that due to the deployment of the RIS, the effective channel $\bm G^{\rm AB}_\ell$ between agent A and agent B can be actively configured in each sensing stage.

Similarly, the received pilot in the $\ell$-th round at agent A is given by 
\begin{align}
    {y}^{\rm A}_\ell = ({\bm{w}}^{\rm A}_{{\rm r},\ell})^{\sf H} {\bm G}_{\ell}^{\rm BA} {\bm{w}}^{\rm B}_{{\rm t},\ell} {x}^{\rm B}_\ell+{n}^{\rm A}_\ell,\quad\ell=0,\cdots,L-1,
\end{align}
where ${\bm G}_{\ell}^{\rm BA} = \bm T\diag({\bm v}^{\rm BA}_\ell)\bm R$ is the cascaded channel from agent B to agent A with the RIS configuration ${\bm v}^{\rm BA}_\ell$, the vectors ${\bm{w}}^{\rm A}_{{\rm r},\ell}\in\mathbb{C}^{M_{\rm t}}$ and ${\bm{w}}^{\rm B}_{{\rm t},\ell}\in\mathbb{C}^{M_{\rm r}}$ are the sensing beamformers at agent A and agent B, respectively, and ${n}^{\rm A}_\ell\sim\mathcal{CN}(0,\sigma^2)$ is the effective additive Gaussian noise.

\subsection{Active Sensing for  Beam and Reflection Alignment}
The ping-pong pilots are used to design the sensing beamformers and RIS
coefficients over the multiple pilot stages and the eventual data transmission
beamformers and reflection.  The beamformer design is the same as in the
first part of the paper, i.e., as in \eqref{eq:pilot_rx}, \eqref{eq:pilot_tx}, and \eqref{eq:final_w}.
Below we focus on the main additional component for the RIS-assisted system, i.e., the need to design the reflection coefficients both in the pilot sensing phase and the data transmission phase.
We assume that both agent A and agent B can communicate with the RIS controller, so the RIS has access to the historical observations from both sides.

As shown in Fig.~\ref{fig:pilot_transmission_pingpong_RIS}, at the beginning of the $\ell$-th ping-pong round, agent A sends a pilot to agent B. After this pilot is received at agent B, the available observations at the RIS controller are $\{y^{\rm A}_i\}_{i=0}^{\ell-1}$ and $\{y^{\rm B}_i\}_{i=0}^{\ell}$. 
The RIS reflection coefficients in the next pilot transmission in the direction from agent B to agent A can now be designed as 
\begin{align}\label{eq:pilot_rx3}
    {\bm v}^{\rm BA}_{\ell} = {f}^{\rm BA}_{\rm v,\ell}\left(\{y^{\rm B}_{i}\}_{i=0}^{\ell}, \{y^{\rm A}_{i}\}_{i=0}^{\ell-1} \right), ~ \ell=0,\cdots,L-1,  
\end{align} 
where ${f}^{\rm BA}_{\rm v,\ell}: \mathbb{C}^{2\ell+1}\rightarrow\mathbb{C}^{N}$ is the adaptive sensing scheme at the RIS, with unit modulus constraint on the entries of the output, i.e., $\left|[{\bm v}^{\rm BA}_{\ell}]_i\right|=1$, for $i=1,\cdots,N$.

After the pilot from the agent B is received at the agent A, both agents now have $\ell+1$ observations,
i.e., $\{y^{\rm A}_i\}_{i=0}^{\ell}$ and $\{y^{\rm B}_i\}_{i=0}^{\ell}$. 
The RIS controller can then design the next reflection coefficients for the direction from agent A to agent B for the $(\ell+1)$-th round as follows: 
\begin{align}\label{eq:pilot_tx3}
    \bm v_{\ell+1}^{\rm AB} = {f}^{\rm AB}_{{\rm v,},\ell}\left(\{y^{\rm A}_i\}_{i=0}^{\ell},\{y^{\rm B}_{i}\}_{i=0}^{\ell} \right),  ~ \ell=0,\cdots, L-2,
\end{align}
where ${f}^{\rm AB}_{\rm v,\ell}: \mathbb{C}^{2(\ell+1)}\rightarrow\mathbb{C}^{N}$ is the active sensing strategy with unit modulus constraint on the entries of the output, i.e., $\left|[\bm v^{\rm AB}_{\ell+1}]_i\right|=1$, for $i=1,\cdots,N$. 
In the initial stage, the reflection coefficients $\bm v^{\rm AB}_0$ are set to be some fixed vector. 

After $L$ rounds of pilot training, 
the RIS controller designs the final reflection coefficients for data transmission as follows:
\begin{equation}\label{eq:final_v}
    {\bm v} = g_{\rm v}\left(\{y^{\rm A}_i\}_{i=0}^{L-1}, \{y^{\rm B}_{i}\}_{i=0}^{L-1}\right),
\end{equation}
where $ g_{\rm v}: \mathbb{C}^{2L}\rightarrow\mathbb{C}^{N}$ maps the received pilots at the Tx and the Rx to the RIS reflection coefficients with unit modulus constraint.

Thus, the overall problem can be formulated as:
\begin{subequations}\label{eq:problem_formulation_RIS}
    \begin{align}
        &\underset{\mathcal{F}}{\Maximize} ~~&&\mathbb{E}\left[|\bm w_{\rm r}^{\sf H}\bm R^{\sf H}\diag(\bm v^{\sf H})\bm T^{\sf H}\bm w_{\rm t}|^2\right]\\
        &\subj ~ &&\eqref{eq:pilot_tx}, \eqref{eq:pilot_tx3}, \eqref{eq:pilot_rx},\eqref{eq:pilot_rx3}, \eqref{eq:final_w} ~\text{and}~\eqref{eq:final_v},
    \end{align}
\end{subequations}
where the optimization variables are a set of functions 
\begin{equation}
    \begin{aligned}
        \mathcal{F}=&\left\{\{f^{\rm A}_{\rm t, \ell}(\cdot)\}_{\ell=0}^{L-2},\{f^{\rm A}_{\rm r,\ell}(\cdot)\}_{\ell=0}^{L-2}, \{f^{\rm AB}_{\rm v, \ell}(\cdot)\}_{\ell=0}^{L-2},g_{\rm t}(\cdot),g_{\rm v}(\cdot)
        \right.\\
        &~\left. \{f^{\rm B}_{\rm t, \ell}(\cdot)\}_{\ell=0}^{L-1},\{f^{\rm B}_{\rm r,\ell}(\cdot)\}_{\ell=0}^{L-2},\{f^{\rm BA}_{\rm v, \ell}(\cdot)\}_{\ell=0}^{L-1},
        g_{\rm r}(\cdot)\right\},
    \end{aligned}
\end{equation}
and the expectation is taken over all the random variables in the system, i.e., the channels and the noise.

In comparison to problem \eqref{eq:problem_formulation}, the problem \eqref{eq:problem_formulation_RIS} is more challenging since it involves designing additional functions, i.e., $\{f^{\rm AB}_{\rm v, \ell}(\cdot)\}_{\ell=0}^{L-2}, \{f^{\rm BA}_{\rm v, \ell}(\cdot)\}_{\ell=0}^{L-1}, g_{\rm v}(\cdot)$ and has more constraints to satisfy. Nonetheless, these additional functions and constraints can be readily included in the proposed deep active sensing framework by using additional fully connected DNNs that can output the designed reflection coefficients with unit modulus constraints.

\subsection{Proposed Deep Learning Framework}
\begin{figure}[t]
    \centering
    \includegraphics[width=7.88cm]{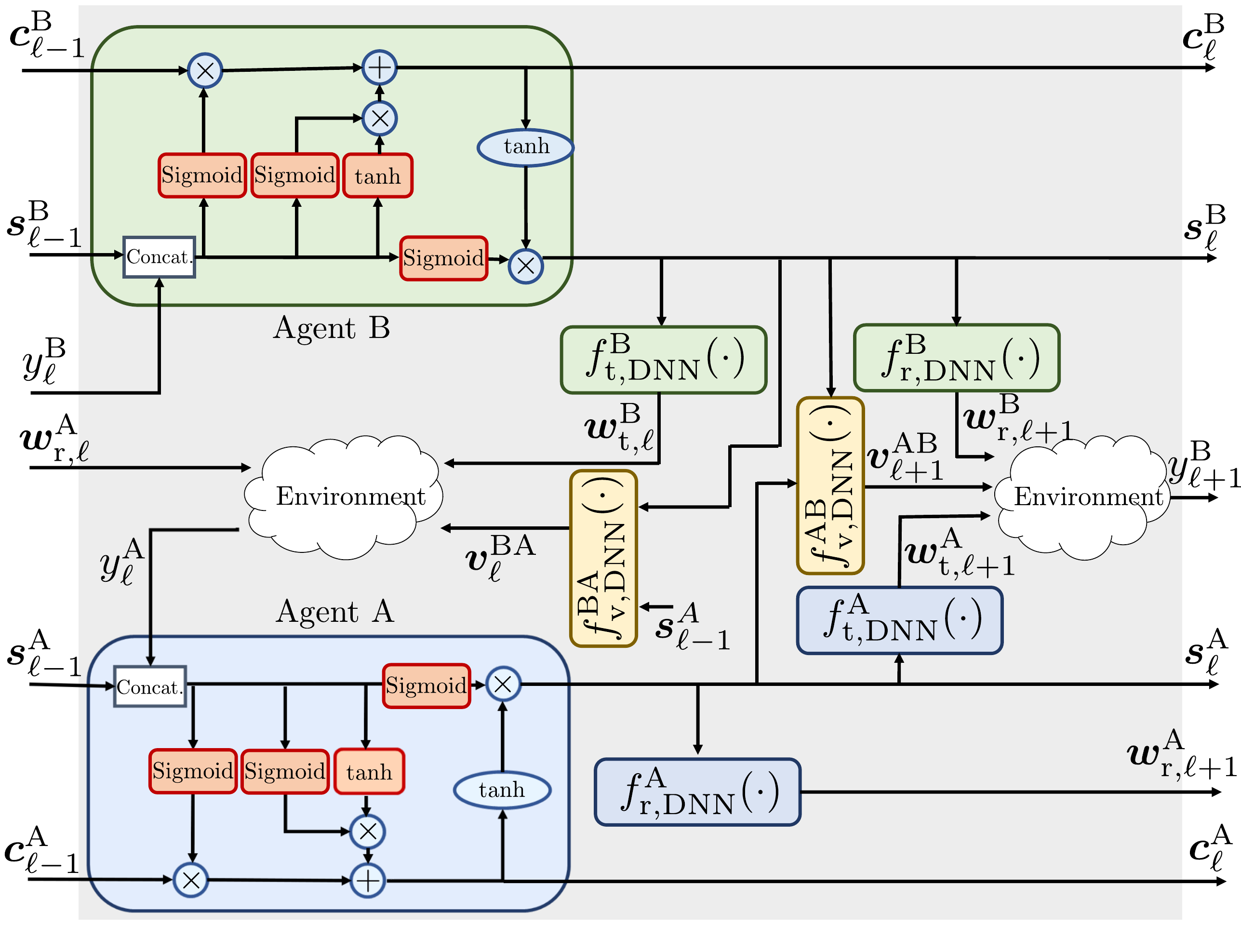}
    \caption{Proposed active sensing unit for the two-sided beam and reflection alignment problem in the $\ell$-th ping-pong pilot training round.}\label{fig:lstm_framework_RIS}
\end{figure}

We extend the proposed active sensing framework to include the RIS controller as shown in Fig.~\ref{fig:lstm_framework_RIS}, which describes the active sensing unit for the $\ell$-th pilot transmission round. The sensing beamformers are designed as in \eqref{eq:pilot_rx} and \eqref{eq:pilot_tx}.
The reflection coefficients $\bm v^{\rm AB}_{\ell+1}$ for the next pilot training in the direction from agent A to agent B are learned by a fully connected neural network with the hidden state vectors ${\bm s}^{\rm A}_{\ell}$ and $\bm{s}^{\rm B}_\ell$ as inputs, i.e.,
\begin{equation}\label{eq:ris_sensing1}
    \bm v^{\rm AB}_{\ell+1} = f_{\rm v, DNN}^{\rm AB} ([({\bm s}^{\rm A}_{\ell})^\top, (\bm{s}^{\rm B}_\ell)^\top]^\top).
\end{equation}
Likewise, in the direction from agent B to agent A, the reflection coefficients ${\bm v}^{\rm BA}_\ell$ are actively designed by a fully connected neural network with ${\bm s}^{\rm A}_{\ell-1}$ and $\bm{s}^{\rm B}_{\ell}$ as inputs, i.e., 
\begin{equation}\label{eq:ris_sensing2}
    {\bm v}^{\rm BA}_\ell = f_{\rm v, DNN}^{\rm BA} ( [(\bm{s}^{\rm A}_{\ell-1})^\top,(\bm s_{\ell}^{\rm B})^\top]^\top).
\end{equation}
To train the neural networks in the active sensing unit, we concatenate $L$ active sensing units together and form a very deep neural network, corresponding to $L$ rounds of pilot transmission as in Fig.~\ref{fig:pilot_transmission_pingpong_RIS}.
The overall deep active sensing architecture is similar to the one in Fig.~\ref{fig:overall_lstm}, except that the active sensing unit is replaced by the one shown in Fig.~\ref{fig:lstm_framework_RIS}. Further, an additional fully connected neural network $g_{{\rm v, DNN}}(\cdot)$ is used to map the final cell state from both sides to the reflection coefficients $\bm v$ for data transmission, 
i.e., 
\begin{align}\label{eq:final_ris}
    \bm{v}  = g_{{\rm v, DNN}} ([(\bm{c}^{\rm A}_{L-1})^\top, ({\bm c}^{\rm B}_{L-1})^\top]^\top).
\end{align}
The final beamforming vectors $\{\bm w_{{\rm t}}, \bm w_{\rm r}\}$ for data transmission are obtained as before, i.e., as in \eqref{eq:final_w}.

\subsection{Simulation Results}
\label{sec:simulation_results}
\begin{figure}[t]
    \centering
    \includegraphics[width=8.6cm]{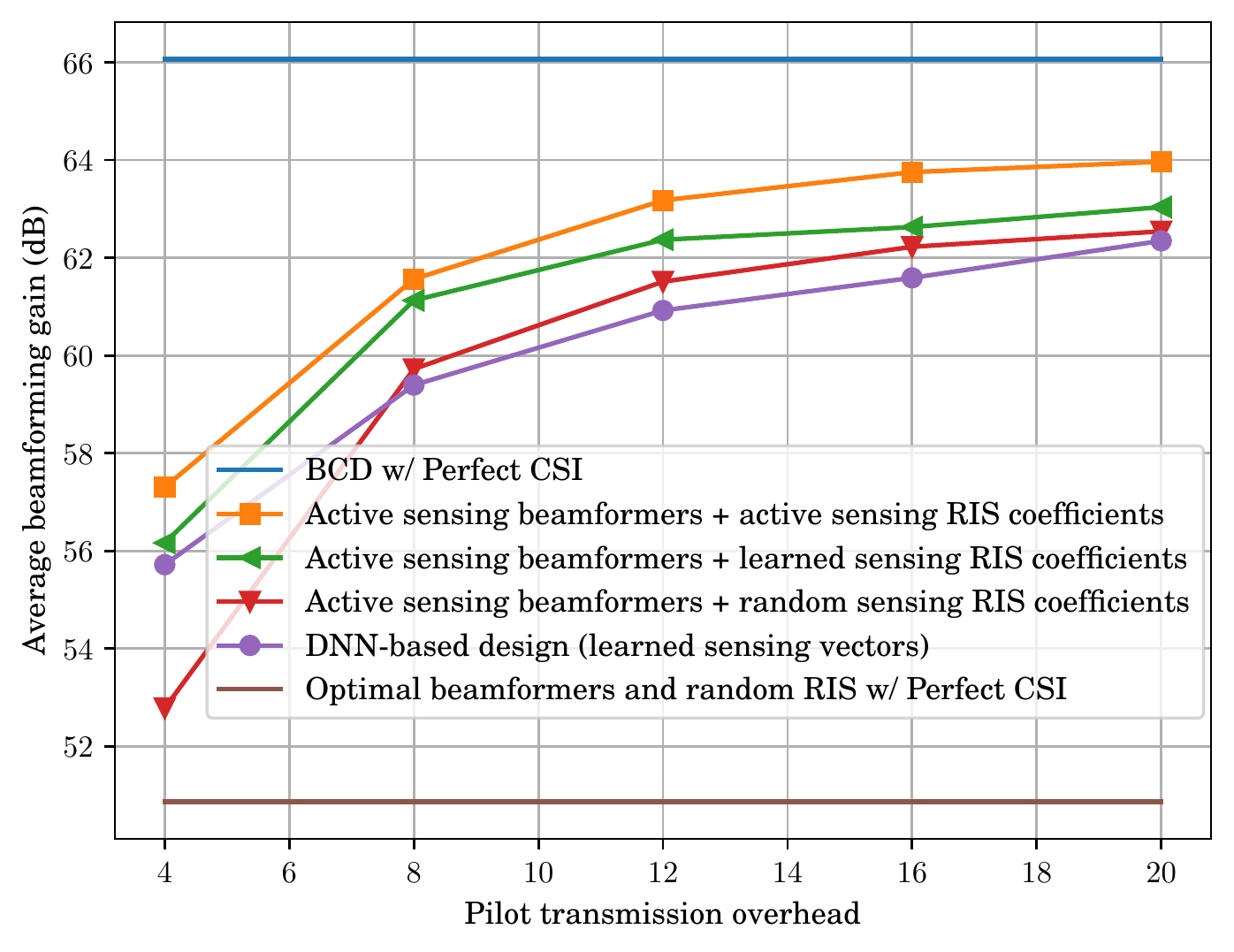}
    \caption{Average beamforming gain vs. pilot overhead in the RIS-assisted system with $M_{\rm t}=64$, $M_{\rm r}=32$, $N=64$, and  $P_1/\sigma^2=P_2/\sigma^2=0$ dB.}\label{fig:bf_gain_RIS}
\end{figure}
\begin{figure*}
    \centering
    \includegraphics[width=18cm]{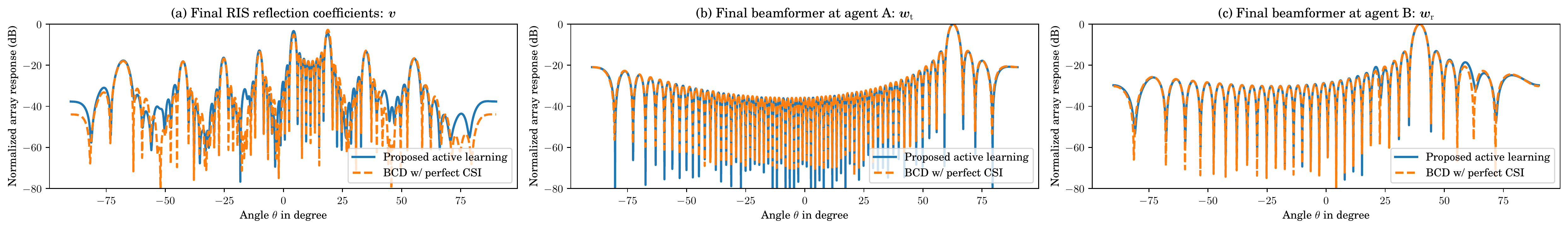}
    \caption{Final beamforming pattern in the RIS-assisted system, with $M_{\rm t}=64$, $M_{\rm r}=32$, $N=64$, $L=8$, and $P_1/\sigma^2=P_2/\sigma^2=0$dB.}\label{fig:bf_final__RIS_pattern}
\end{figure*}
In this subsection, we illustrate the performance of the proposed deep active sensing framework for the two-sided beam and reflection alignment problem. We consider an RIS-assisted mmWave system where there are $8\times 8$ elements at the RIS. The other simulation parameters are the same as in Section \ref{sec:sim_hybrid}. For the implementation of active sensing framework, we follow the settings in \ref{sec:impl_hybrid}. The fully connected neural networks in \eqref{eq:ris_sensing1}, \eqref{eq:ris_sensing2} and \eqref{eq:final_ris} are respectively of sizes $[512,512,2M_{\rm t}]$, $[512,512,2M_{\rm t}]$, and $[1024,1024,2N]$. 

We compare with the following benchmarks:

\emph{BCD with perfect CSI:} Given perfect CSI, the beamforming gain $|\bm w_{\rm r}^{\sf H}\bm R^{\sf H}\diag(\bm v^{\sf H})\bm T^{\sf H}\bm w_{\rm t}|^2$ can be optimized by the block coordinate descent (BCD) algorithm \cite{9110912}. That is, beamformers $\{\bm w_{\rm t}, \bm w_{\rm r} \}$ and reflection coefficients $\bm v$ are optimized alternately until the algorithm converges. In particular, given the reflection coefficients $\bm v$, the optimal beamformers $\{\bm w_{\rm t}, \bm w_{\rm r} \}$ can be found by the SVD method in \eqref{eq:optimal_bf}; given the beamformers $\{\bm w_{\rm t}, \bm w_{\rm r} \}$, the optimal reflection coefficients can be found via phase matching \cite{9110912}.

\emph{DNN-based design (learned sensing vectors):} We follow the ping-pong pilot training protocol in Fig.~\ref{fig:pilot_transmission_pingpong_RIS}, but the sensing beamformers and reflection coefficients are learned from the channel statistics. This is done by setting the sensing vectors as trainable parameters in the neural network training phase, so they are updated by the SGD algorithm according to the training data, but are fixed during testing. The received pilots at agent A and agent B are mapped to the corresponding beamformers for data transmission by two different DNNs of sizes $[1024,1024,2M_{\rm t}]$ and $[1024,1024,2M_{\rm r}]$, respectively. The reflection coefficients for data transmission are obtained by using a DNN of size $[1024,1024,2N]$; the DNN takes the received pilots from both sides as inputs. 

\emph{Optimal beamformer and random RIS with perfect CSI:} To see the benefits of optimizing the RIS, we also use a baseline with randomly generated RIS coefficients, but the beamformers are optimized assuming perfect CSI using SVD as in \eqref{eq:optimal_bf}.

For the proposed active sensing approach, the sensing beamformers of both agents are always actively
designed, but we test different cases with the reflection coefficients being random, or
learned from channel statistics, or actively designed. In Fig.~\ref{fig:bf_gain_RIS}, we plot
the average beamforming gain over $10000$ channel realizations against the
pilot training overhead. 
It can be seen that the actively designed beamforming and RIS reflective coefficients
achieve the best performance among all learning strategies. 
In particular, the performance gain over the randomly designed
RIS with active beamformers is around $2$dB and over the learned sensing RIS
coefficients with active beamformers is around $1$dB. These results show the advantage 
of actively designing the RIS reflection coefficients in the sensing phase.      

\subsection{Interpretation of Solutions from Active Sensing}

To understand the solution learned by the active sensing framework, we plot the normalized beamforming pattern of the beamformers and the RIS coefficients. The normalized array response of the beamformers are given by \eqref{eq:bf_gain_effective}. The normalized array response  of the RIS with coefficients $\bm v\in\mathbb{C}^N$ is given as 
\begin{align}\label{eq:bf_gain_effective_RIS}
    \tilde{f}_{\rm beam}(\bm v,\theta) = \frac{1}{N}|\bm v^{\sf H}\bm a(\theta)|^2,
\end{align} 
where $\bm a(\theta)=\frac{1}{\sqrt{N}} [1,\cdots,e^{j\pi(N-1)\sin(\theta)}]^\top$ is the normalized steering vector.

In Fig.~\ref{fig:bf_final__RIS_pattern}, we compare the beamforming pattern of the final learned beamformers and RIS coefficients with the ones designed by the BCD method with perfect CSI. The number of ping-pong rounds $L=8$. It can be seen that the beamforming patterns given by the proposed active sensing method perfectly match the ones assuming perfect CSI. This verifies that the proposed active sensing method indeed outputs meaningful solutions.

\section{Conclusions}\label{sec:conclusion}

This paper proposes an active sensing framework to sequentially design the
sensing vectors from successive observations for the two-sided beam alignment
problem in mmWave systems.  The proposed approach is based on a novel ping-pong
pilot training scheme, which eliminates the need for feedback between the Tx
and the Rx at the sensing stage, and is data driven, so that it does not rely
on mathematical models of the channel.  Active sensing is a challenging
problem because of the sequential nature of the sensing operation, which allows
successive exploration of the channel landscape, but also requires the learning
agent to succinctly summarize the observations so far, which is a nontrivial
task.  To this end, this paper proposes to use an active sensing strategy
based on LSTM neural networks to account for the sequential dependency of the
sensing task. 
The proposed active sensing framework can be extended to the RIS-assisted
mmWave system, in which the RIS can also be actively configured in the pilot
training phase. Simulation results verify the superior performance of the
proposed method as compared to the previous state-of-the-art methods and that 
the learned beamforming and reflection patterns are meaningful and interpretable.

\bibliographystyle{IEEEtran}
\bibliography{IEEEabrv,refs}
\end{document}